\pdfoutput=1
\RequirePackage{ifpdf}
\ifpdf 
\documentclass[pdftex]{sigma}
\else
\documentclass{sigma}
\fi

\numberwithin{equation}{section}

\newtheorem{Theorem}{Theorem}[section]
\newtheorem{Proposition}[Theorem]{Proposition}
{\theoremstyle{definition}
\newtheorem{Remark}[Theorem]{Remark}}

\begin{document}

\allowdisplaybreaks

\renewcommand{\thefootnote}{$\star$}

\newcommand{\arXivNumber}{1603.03570}

\renewcommand{\PaperNumber}{073}

\FirstPageHeading

\ShortArticleName{Large $N$ Limits in Tensor Models}

\ArticleName{Large $\boldsymbol{N}$ Limits in Tensor Models: Towards More\\
Universality Classes of Colored Triangulations\\
in Dimension $\boldsymbol{d\geq 2}$\footnote{This paper is a~contribution to the Special Issue on Tensor Models, Formalism and Applications. The full collection is available at \href{http://www.emis.de/journals/SIGMA/Tensor_Models.html}{http://www.emis.de/journals/SIGMA/Tensor\_{}Models.html}}}

\Author{Valentin BONZOM}

\AuthorNameForHeading{V.~Bonzom}

\Address{LIPN, UMR CNRS 7030, Institut Galil\'ee, Universit\'e Paris 13,\\
Sorbonne Paris Cit\'e, 99, avenue Jean-Baptiste Cl\'ement, 93430 Villetaneuse, France}
\Email{\href{mailto:bonzom@lipn.univ-paris13.fr}{bonzom@lipn.univ-paris13.fr}}
\URLaddress{\url{http://lipn.univ-paris13.fr/~bonzom/}}

\ArticleDates{Received March 14, 2016, in f\/inal form July 20, 2016; Published online July 24, 2016}

\Abstract{We review an approach which aims at studying discrete (pseudo-)manifolds in dimension $d\geq 2$ and called random tensor models. More specif\/ically, we insist on genera\-li\-zing the two-dimensional notion of $p$-angulations to higher dimensions. To do so, we consider families of triangulations built out of simplices with colored faces. Those simplices can be glued to form new building blocks, called bubbles which are pseudo-manifolds with boun\-da\-ries. Bubbles can in turn be glued together to form triangulations. The main challenge is to classify the triangulations built from a given set of bubbles with respect to their numbers of bubbles and simplices of codimension two. While the colored triangulations which maximize the number of simplices of codimension two at f\/ixed number of simplices are series-parallel objects called melonic triangulations, this is not always true anymore when restricting attention to colored triangulations built from specif\/ic bubbles. This opens up the possibility of new universality classes of colored triangulations.
We present three existing strategies to f\/ind those universality classes. The f\/irst two strategies consist in building new bubbles from old ones for which the problem can be solved. The third strategy is a bijection between those colored triangulations and stuf\/fed, edge-colored maps, which are some sort of hypermaps whose hyperedges are replaced with edge-colored maps. We then show that the present approach can lead to enumeration results and identif\/ication of universality classes, by working out the example of quartic tensor models. They feature a tree-like phase, a~planar phase similar to two-dimensional quantum gravity and a phase transition between them which is interpreted as a proliferation of baby universes.
While this work is written in the context of random tensors, it is almost exclusively of combinatorial nature and we hope it is accessible to interested readers who are not familiar with random matrices, tensors and quantum f\/ield theory.}


\Classification{05C10; 05C75; 83C45; 81T18; 83C27}

\renewcommand{\thefootnote}{\arabic{footnote}}
\setcounter{footnote}{0}

\section{Introduction}

\subsection{Random matrices and combinatorial maps}

Random matrix models are probability distributions for a random (typically Hermitian) mat\-rix~$A$ of size $N\times N$. It takes the form $\exp\{ -N \operatorname{Tr} V(A)\}$ where $V$ is typically a polynomial~\cite{matrix}. Expectations of unitary invariant polynomials like $\operatorname{Tr} A^n$ are in fact generating functions for connected combinatorial maps rooted on an edge incident to a vertex of degree $n$ (also known as ribbon graphs in physics). We will be interested in generalizing this approach to higher-dimensional objects, and more particularly in generalizing the $2p$-angulations which are combinatorial maps whose faces all have degree~$2p$.

A connected combinatorial map $M$ is a connected graph\footnote{Multiple edges and loops are allowed.} equipped with a cyclic order of the edges incident to each vertex. In other words, each vertex is locally embedded in a small disc. A corner is the portion of the disc between two consecutive edges at a vertex and we orient them counter-clockwise from one edge to the other. A~face of~$M$ is a cycle formed by following edges and corners counter-clockwise. The genus~$g(M)$ of the map satisf\/ies $2 - 2g(M) = F(M) - E(M) + V(M)$ where $F(M)$, $E(M)$, $V(M)$ are the number of faces, edges and vertices of~$M$, respectively. Topologically, this genus is the smallest value of the genus of a surface on which~$M$ can be embedded without crossings. Combinatorial maps thus def\/ine a notion of discrete surfaces which generalize triangulations (a triangulation being a combinatorial map in which all faces have degree three).

Let us consider the following example of a random matrix model. Def\/ine the partition function
\begin{gather} \label{2pMatrixModel}
Z(N,t) = \int dA\, \exp\big\{{-}N\bigl( \operatorname{Tr} A^2 + t \operatorname{Tr} A^{2p}\bigr)\big\},
\end{gather}
and free energy $f(N, t) = \ln Z(N, t)$. Then $\frac{1}{Z(N, t)}\exp\big\{{-}N\bigl( \operatorname{Tr} A^2 + t \operatorname{Tr} A^{2p}\bigr)\big\}$ def\/ines a probability distribution, and the so-called $n$-point functions are the following expectations
\begin{gather}
\big\langle \operatorname{Tr} A^n \big\rangle = \frac{1}{Z(N, t)} \int dA \operatorname{Tr} (A^n) \exp\big\{{-}N\big( \operatorname{Tr} A^2 + t \operatorname{Tr} A^{2p}\big)\big\}.
\end{gather}
To unravel the connection with combinatorial maps, one expands $e^{-N t \operatorname{Tr} A^{2p}}$ and (illegally) commutes the sum with the integral. For the free energy, one gets
\begin{gather}
f(N, t) = \ln \sum_{k\geq 0} \frac{(-N t)^k}{k!} \int dA \underbrace{\operatorname{Tr}\big(A^{2p}\big) \cdots \operatorname{Tr}\big(A^{2p}\big)}_{\text{$k$ times}}\, e^{-N \operatorname{Tr} A^2}.
\end{gather}
It is thus suf\/f\/icient to evaluate the expectation of the product $(\operatorname{Tr} (A^{2p}))^k$ in the Gaussian distribution. According to Wick's theorem, such an expectation is the sum over all pairings (i.e., perfect matchings) of the $2p\times k$ copies of $A$, weighted in a certain way (which we will not explain here). One represents each~$A^{2p}$ as a vertex of degree~$2p$, and a pairing consists in drawing an edge between two vertices when some of the $A$s are paired. A careful inspection of Wick's theorem in the case of matrix variables reveals that the cyclic order of the edges around each vertex \emph{does} matter. That is, it is an expansion onto combinatorial maps with~$k$ vertices of degree~$2p$. Taking the logarithm then restricts the sum over connected maps.

Notice that this correspondence between matrix models and generating functions of maps is obtained by considering matrix integrals as formal power series, here in the parameter $t$. The matrix integral itself is however only def\/ined for $t$ with a positive real part. Establishing a~rigorous relationship between the formal power series and the integrals is the purpose of constructive f\/ield theory; we refer to~\cite{GurauKrajewski} for an explicit and state-of-the-art example in the matrix case.

Let ${\mathcal M}_{2p}$ be the set of connected maps with vertices of degree $2p$. The free energy then expands as
\begin{gather} \label{MapsFreeEnergy}
f(N, t) = \sum_{M\in {\mathcal M}_{2p}} \frac{(2p)^{V(M)}}{s(M)} N^{2 - 2g(M)} (-t)^{V(M)},
\end{gather}
where $V(M)$ is the number of vertices and $s(M)$ a combinatorial factor related to the symmetries of~$M$. Therefore, random matrix models provide generating functions of maps with prescribed vertex degrees, counted with respect to their genera and numbers of vertices. Applying the same reasoning to the expectation $\langle \operatorname{Tr} A^n\rangle$ leads to a sum over rooted maps, i.e., maps with a~distinguished oriented edge outgoing at a vertex of degree~$n$.

\subsection[Colored triangulations with prescribed bubbles as generalizations of $p$-angulations]{Colored triangulations with prescribed bubbles\\ as generalizations of $\boldsymbol{p}$-angulations}

A $p$-angulation is a map in which all faces have degree (i.e., length) $p$. Duality is an operation which from a map $M$ produces another map $D(M)$ in the following way. Each face $f$ of $M$ is represented as a~vertex~$f^*$ of the dual map~$D(M)$. When two faces $f_1$, $f_2$ share an edge in~$M$, there is a dual edge in $D(M)$ which connects $f_1^*$ and $f_2^*$. The order around a vertex $f^*$ is inherited from the cyclic order in which the edges are encountered along~$f$. Duality thus turns $p$-angulations into maps with vertices of degree $p$ bijectively. Therefore one can consider the matrix model~\eqref{2pMatrixModel} as a generating function of $2p$-angulations.

By generalizing the above approach to higher dimensions we mean introducing a family of higher-dimensional discrete manifolds (e.g., simplicial complexes) and their generating function as in \eqref{MapsFreeEnergy}. Any attempts to do so then face the following challenges.
\begin{itemize}\itemsep=0pt
\item Find an equivalent in some sense to the genus of a map (and we know topology in dimension higher than two is not characterized by a single number). Maps with vertex degrees equal to $2p$ satisfy $E(M) = p V(M)$, which simplif\/ies Euler's relation to
\begin{gather*}
2 - 2g(M) = F(M) - (p-1) V(M) \leq 2.
\end{gather*}
The fact that $g(M) \geq 0$ means that there is a bound on the number of faces which is linear in the number of vertices. This is precisely the kind of equations we would like to (and will) extend to higher dimensions.
\item Is there a generalization of random matrices to other random objects which generate a~family of higher-dimensional discrete manifolds?
\end{itemize}

The answer to the second point is that \emph{random tensors} generate discrete (pseudo-)manifolds of dimensions higher than 2 by means of the same mechanism as the one from \eqref{2pMatrixModel}--\eqref{MapsFreeEnergy}, as shown in \cite{AmbjornTensors, GrossTensors, SasakuraTensors} all coming up in the early 90s. However, the methods used in most matrix models, e.g., reduction to eigenvalues and orthogonal polynomials, are not available to random tensors. It means that the analysis of those models has to rely on \emph{combinatorial methods}. This was a problem because those early attempts generate wild objects, which may not be ``nice'' cellular decompositions of pseudo-manifolds \cite{Lost,Lost++,Lost+}, and whose combinatorics is therefore too dif\/f\/icult to control.

The dif\/f\/iculty to control both the topology and the combinatorics of the objects generated by tensor models explains that they lay dormant for twenty years. It was eventually realized that \emph{colored triangulations} form a family of higher-dimensional triangulations which may be suitable to combinatorics. At least, they were designed in the context of topology~\cite{FerriGagliardi, Lins}, which solve part of the topological issues. It was shown that colored triangulations are piecewise-linear pseudo-manifolds and all piecewise-linear manifolds admit colored triangulations. We refer to the review~\cite{GurauRyanReview} for details on the structures of colored triangulations, in particular in the context of tensor models.

\begin{figure}[t]
\centering\includegraphics[scale=1]{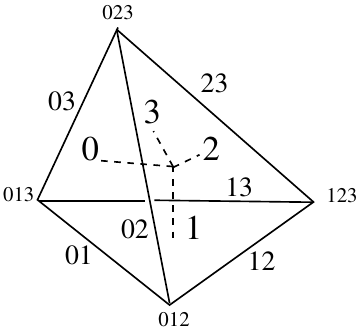}
\caption{This is a colored tetrahedron with faces colored $0$, $1$, $2$, $3$. Edges and vertices are respectively identif\/ied by pairs and triples of colors.}\label{fig:ColoredTet}
\end{figure}

Colored triangulations are gluings of colored simplices. A colored $d$-simplex is an abstract simplex of dimension $d$ with its $d+1$ faces (i.e., boundary $(d-1)$-simplices) colored from~$0$ to~$d$ as shown in Fig.~\ref{fig:ColoredTet}. The face-coloring induces colorings on the subsimplices. Indeed, a~$(d-2)$-simplex is at the intersection of exactly two $(d-1)$-simplices and is therefore identif\/ied by the couple of colors of those two $(d-1)$-simplices. Similarly $(d-3)$-simplices are identif\/ied by triplets of colors and so on. In 3D for instance, the colors are $0$, $1$, $2$, $3$ on the four triangles of a~tetrahedron. An edge is identif\/ied by a couple of colors (those of the two triangles which meet at that edge) and vertices are identif\/ied by triplets of colors.

\begin{figure}[t]
\centering\includegraphics[scale=1]{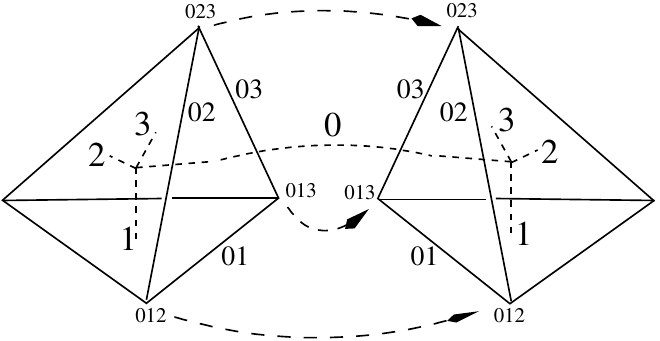}
\caption{The rule of colored gluing is one glues two tetrahedra of opposite orientations along a face such that all induced colorings of the sub-simplices are preserved.}\label{fig:GluingTet}
\end{figure}

Those induced colorings provide a canonical gluing rule. Two $d$-simplices have to be glued by identifying two of their $(d-1)$-simplices. There are generally lots of ways to do so, but there is a unique way which respects all the induced colorings. In~$3D$, that would mean identifying a~triangle of color $c\in\{0, 1, 2, 3\}$ with another one, such that the edge of colors $(c, c')$ for $c'\neq c$ of one tetrahedron is identif\/ied with the edge of the same colors on the other tetrahedron and similarly for the three vertices of the triangle. This is shown in Fig.~\ref{fig:GluingTet}.

The combinatorial upside is that colored triangulations can be represented using \emph{colored graphs}, i.e., regular graphs with colored edges. Indeed, each simplex is represented by a vertex of the graph and there is an edge of color $c\in \{0, \dotsc, d\}$ between two vertices of the graph if the two corresponding $d$-simplices are glued along a $(d-1)$-simplex of color $c$. All the vertices have degree $d+1$, which is the number of $(d-1)$-simplices around a simplex, and all edges incident at a vertex have distinct colors. A $(d-2)$-simplex on the boundary of a $d$-simplex is identif\/ied by a pair of colors. When that $(d-2)$-simplex is shared with other $d$-simplices, it is still labeled by the same pair of colors and it is identif\/ied in the colored graph by a path alternating the two colors. When the $(d-2)$-simplex, with colors $(c c')$, is not on the boundary of the gluing, it is then represented in the graph as a cycle alternating the colors $c$ and~$c'$. We call those cycles \emph{faces} and they represent simplices of codimension two.

There is a tensor model which generates colored triangulations (or equivalently colored graphs) in any f\/ixed dimension $d\geq 2$ \cite{ColoredMelonic}. Gurau was able to study its colored graphs combinatorially and proved this way that there is a linear bound on the number of faces. The distance to the upper bound is measured by Gurau's degree, which thus extends the genus of maps. In quantum f\/ield theory terminology it means that the tensor model admits a $1/N$-expansion, i.e., the free energy is bounded like $f \leq N^D$ for some $D$, \cite{LargeN1, LargeN3, LargeN2}. At $d\geq 3$, the graphs which dominate at large $N$ are those which generalize the genus zero maps from $d=2$. They are those which maximize the number of faces at f\/ixed number of vertices, just like at $d=2$. For $d\geq 3$, it was shown that they are series-parallel graphs called \emph{melonic} graphs. They are in bijection with trees and their enumeration is straightforward~\cite{ColoredMelonic}. Next-to-leading-order graphs were discovered in~\cite{NLO} by Kaminski, Oriti and Ryan, and eventually the complete classif\/ication of colored graphs was performed with respect to Gurau's degree in~\cite{GurauSchaeffer} by Gurau and Schaef\/fer.

\begin{figure}[t]
\centering\includegraphics[scale=.5]{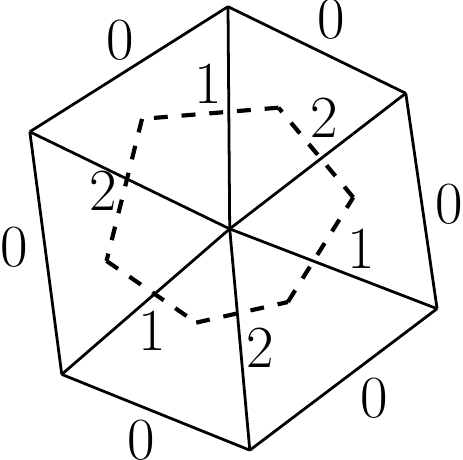}
\caption{A $2p$-angle, drawn in solid lines, with boundary edges of color $0$ can be obtained as the gluing of~$2p$ colored triangles. The dashed lines represent the dual object with all the colors except~$0$, with a~vertex for each triangle and an edge of color~$c$ between two vertices if the two triangles are glued along an edge of color~$c$.}\label{fig:2pAngle}
\end{figure}

In contrast with \cite{GurauSchaeffer} in which the set of all colored graphs is considered, we wish to study generalizations of $2p$-angulations which only correspond to a subset of colored graphs. A $2p$-angulation is a gluing of $2p$-angles, each of which can be seen as a gluing of $2p$ triangles by adding a vertex in the center of each $2p$-angle. It can be done using colored triangles, so that the edges of color $0$ form the boundary of the $2p$-angle, see Fig.~\ref{fig:2pAngle}.

In higher dimensions, a \emph{bubble} is a gluing of colored $d$-simplices whose boundary consists of all $(d-1)$-simplices of color $0$. The choice of the color $0$ is arbitrary, the idea begin that all $(d-1)$-simplices are shared between two simplices except for those of a given color which then form the boundary. The topology of such a gluing is always a disc in two dimensions, but it is typically not a ball in higher dimensions, hence the denomination of bubble instead.

\begin{figure}[t]
\centering\includegraphics[scale=.6]{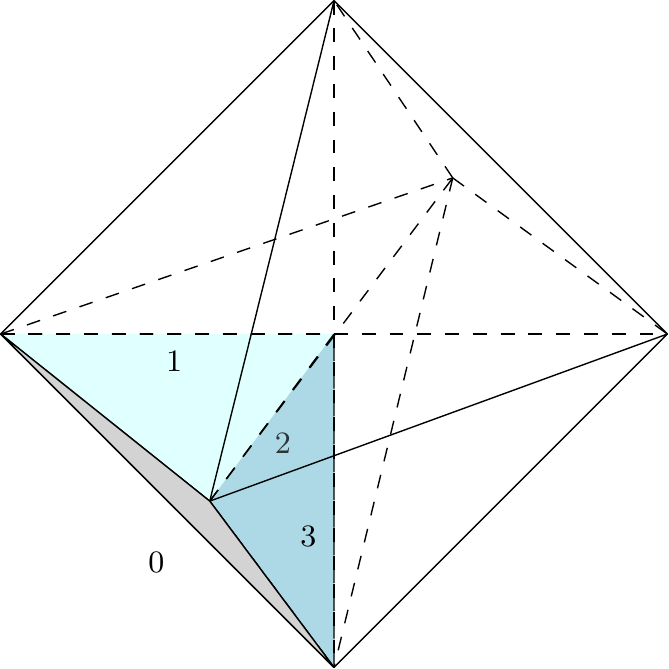} \qquad\qquad \includegraphics[scale=.75]{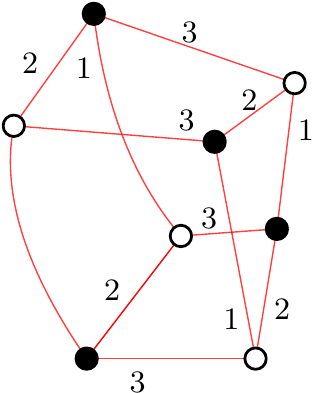}
\caption{On the left is a three-dimensional bubble: a colored bi-pyramid, consisting of 8 tetrahedra, four forming an upward pyramid, four others forming a downward pyramid, and the two pyramids glued to each other, such that all triangles on the boundary have color~$0$. On the right is its the dual colored graph: a vertex for each tetrahedron and an edge of color $c=1, 2, 3$ when two tetrahedra are glued along a triangle of color~$c$.}\label{fig:Bipyramid}
\end{figure}

A bubble can then be represented as a connected colored graph with all the colors except~0, as shown in Fig.~\ref{fig:Bipyramid}. The generalization of the notion of $2p$-angulation which we propose to study consists in choosing a~bubble~$B$, and constructing all colored graphs whose bubbles (i.e., maximal subgraphs with all colors but~$0$) are copies of~$B$. This means taking some copies of $B$ and adding the color $0$ to all vertices so as to get a connected graph. We denote this set of colored graphs ${\mathcal G}(B)$. Interestingly, this set is the one generated by random tensor models in a natural way~\cite{Uncoloring}. When going from a random matrix with distribution of the form $\exp\{-N \operatorname{Tr} V(A)\}$ for a~polynomial~$V$ to a random tensor~$T$ with a distribution of the form $\exp\{-\Phi(T)\}$, it is necessary to specify the admissible forms of~$\Phi$. It turns out that restricting oneself to polynomials invariant under the natural action of $d$ copies of the unitary group is equivalent (via the same steps as~\eqref{2pMatrixModel}--\eqref{MapsFreeEnergy}) to considering the generating functions of colored graphs whose bubbles are prescribed by the choice of $\Phi$. In particular there is a $\Phi_B$ such that
\begin{gather*}
\ln \int dT\,d\overline{T} \, \exp\big\{{-}N^{d-1} \Phi_B(T, \overline{T} ; t)\big\} = \sum_{G\in{\mathcal G}(B)} C(G) N^{d - \omega(G)} t^{b(G)},
\end{gather*}
where we use a complex random tensor with $d$ indices ranging from $1$ to $N$ and its complex conjugate $\overline{T}$. Here $\omega(G) \geq 0$ is Gurau's degree of $G$ and $t$ counts the graphs with respect to the number $b(G)$ of copies of $B$ in $G$ and $C(G)$ is a numerical factor related to the symmetries of~$G$.

On one hand, in order to saturate Gurau's degree, i.e., $\omega(G) = 0$, the allowed bubbles~$B$ must be of the melonic type. This is very restrictive and leads to branched-polymer (i.e., tree-like) behaviors~\cite{Uncoloring, BP}. This suggests that in order to f\/ind more interesting behaviors, one has to study graphs of non-vanishing Gurau's degree, built from non-melonic bubbles. However, a~naive application of the Gurau--Schaef\/fer classif\/ication~\cite{GurauSchaeffer} suggests that for most choices of~$B$, there is always a f\/inite number of graphs of~${\mathcal G}(B)$ at f\/ixed Gurau's degree. This implies that there is no ``large number of bubbles'' limit at f\/ixed Gurau's degree.

It means that if one insists on exploring tensor models and colored triangulations using the generalization of $2p$-angulations provided by the set ${\mathcal G}(B)$, this is a whole new combinatorial analysis which has to be performed. The key point is to modify Gurau's degree and def\/ine a~$B$-degree which depends on the choice of~$B$, such that there are inf\/initely many graphs at f\/ixed degree or at least in the large~$N$ limit (minimal $B$-degree). When this is achieved for a~bubble~$B$, we say that it admits an \emph{enhancement}.

Then the generating function of the graphs with minimal degree can be studied. Its singularity corresponds to the continuum (thermodynamical) limit of the model where the number of bubbles becomes arbitrarily large. It typically behaves like $(t_c - t)^{2-\gamma}$ where $\gamma$ is the entropy exponent. It is the critical exponent which (partly) characterizes the universality class of the continuum limit.

The entropy exponent of melonic graphs is $\gamma_{\text{melons}} = 1/2$, typical of trees, and it is $\gamma_{\text{planar}} = -1/2$ for random planar maps. We will see that tensor models equipped with dif\/ferent bubbles can reproduce both those behaviors as well as the proliferation of baby universes observed in multi-trace matrix models \cite{AlvarezBarbon, BarbonDemeterfi,Das, KlebanovHashimoto, Korchemsky} with $\gamma = 1/3$ (and all the associated multi-critical behaviors which we will not further mention). This will be done very naturally with the ``simplest'' bubbles (those on four vertices).

The present article is mostly a review article. A few new results are included, in Sec\-tions~\ref{sec:BubbleGluing},~\ref{sec:Slices} and~\ref{sec:Enumeration} which in fact extend ideas and results which have already appeared in the literature in specif\/ic cases or more narrow situations than here. We here formalize them and put them in the broader context of enhancements. In the case of Section~\ref{sec:Enumeration}, the new enumeration results generalize~\cite{MelonoPlanar} and remarkably require a totally dif\/ferent approach than there to do so.

The organization is as follows. In Section \ref{sec:Tensors} we review the framework of tensor models and its connection to bubbles and colored graphs. Gurau's degree theorem is given in Section~\ref{sec:Degree}. While it reduces to standard results on combinatorial maps for $d=2$, we explain Gurau's universality theorem for large random tensors~\cite{Universality} and its application to the large $N$ enumeration of melonic graphs for $d\geq 3$~\cite{Uncoloring}. We then start the exploration of other large $N$ limits for tensor models in Section~\ref{sec:New1/N}. We def\/ine enhancements and formulate the combinatorial problem necessary to f\/ind enhancements: f\/ind the colored graphs in~${\mathcal G}(B)$ which maximize the number of faces at f\/ixed number of bubbles. Three strategies are then provided to f\/ind enhancements. The f\/irst two of them consist in f\/inding new enhancements from existing ones, while the third strategy is a true new combinatorial approach to the issue. Those strategies are the following three.
\begin{itemize}\itemsep=0pt
\item New bubbles, with new enhancements, can be created by gluing bubbles whose enhancements are known. We call this inherited enhancements. This gluing of bubbles increases the number of vertices of the bubbles. It was used in~\cite{DoubleScaling} and~\cite{MelonoPlanar} to get results for bubbles with more than four vertices (thus going beyond the quartic tensor models). But it had not been used further than in these specif\/ic cases. This idea had not been synthesized and framed as a strategy to f\/ind enhancements yet. We do so in Section~\ref{sec:BubbleGluing}. It thus contains a~few new equations which extend and synthesize for the f\/irst time techniques used~\cite{DoubleScaling} and~\cite{MelonoPlanar}.
\item The strategy presented in Section~\ref{sec:Slices} consists in looking at sub-bubbles forming a partition of a bubble. If the enhancements of the sub-bubbles are known, the enhancement of the bubble itself can be found. This is equivalent to thinking of colored graphs in dimension~$d$ as superpositions of subgraphs with dimensions~$\Delta_i$, such that $\sum_i \Delta_i = d$. It is a~(new but very modest) extension of the $1/N$-expansions of~\cite{New1/N} to non-trivial enhancements.

\item The third strategy is a bijection, presented in Section~\ref{sec:StuffedMaps} which reviews material from~\cite{StuffedWalshMaps}. Since the challenge is to study faces of colored graphs, it would be enlightening to f\/ind a~bijection with maps for which faces are well-controlled. This would also help understand the complexity of combinatorics on objects of dimension $d\geq 3$ with respect to $d=2$. We present in Section~\ref{sec:StuffedMaps} the bijection introduced in \cite{StuffedWalshMaps} which maps the set ${\mathcal G}(B)$ to some extension of edge-colored hypermaps where hyperedges are replaced with a~``stuf\/f\/ing'', i.e., a~prescribed edge-colored map. This bijection has made it possible to f\/ind enhancements and perform the enumeration of the graphs which maximize the number of faces beyond the case of bubbles with four vertices and beyond melonic bubbles.
\end{itemize}
Finally, we show in Section \ref{sec:Quartic} that the bijection enables to perform enumeration. We do so on the quartic models to keep things simple and also because in that case there are several coupling constants to play with and we can have an explicit algebraic description of the generating function. We f\/ind in particular the exponents $\gamma = 1/2, 1/3, -1/2$ very naturally. Section~\ref{sec:QuarticDominant} reproduces the results of~\cite{MelonoPlanar} using the approach of~\cite{StuffedWalshMaps}. Section~\ref{sec:Enumeration} contains new results as it extends the enumeration performed in~\cite{MelonoPlanar} to additional parameters.

Although the presentation highlights the point of view of random tensor models, the main technique is combinatorics. In fact random tensors and matrices are barely used beyond Section~\ref{sec:Tensors}. We even hope that readers who are unfamiliar with tensor models and quantum f\/ield theory methods can accept the results coming from those f\/ields and go through to the purely combinatorial parts.

\section{Random tensor models} \label{sec:Tensors}

The framework is the one introduced in \cite{Uncoloring}, with an additional freedom on the $N$-dependence.

\subsection{Bubbles and tensorial invariants} \label{sec:Bubbles}

In matrix models, both the potential of the model and the observables are (typically) unitary invariant quantities, e.g.,~$\operatorname{Tr} (M M^\dagger)^p$. To generalize this, we introduce tensorial unitary inva\-riants, or simply inva\-riants. An {\it invariant} is a polynomial in the tensor entries $T_{a_1 \dotsb a_d}$ and $\overline{T}_{a_1 \dotsb a_d}$ which is invariant under the following action of ${\rm U}(N)^d$,{\samepage
\begin{gather*}
T_{a_1 \dotsb a_d} \mapsto \sum_{b_1, \dots, b_d} U^{(1)}_{a_1 b_1} \dotsm U^{(d)}_{a_d b_d} \,T_{b_1 \dotsb b_d},\qquad
\overline{T}_{a_1 \dotsb a_d} \mapsto \sum_{b_1, \dots, b_d} \overline{U^{(1)}}_{a_1 b_1} \dotsm \overline{U^{(d)}}_{a_d b_d} T_{b_1 \dotsb b_d},
\end{gather*}
where $U^{(1)}, \dots, U^{(d)}$ are $d$ independent unitary matrices.}

\begin{figure}[t]\centering
\begin{minipage}[b]{45mm}\centering
\includegraphics[scale=.6]{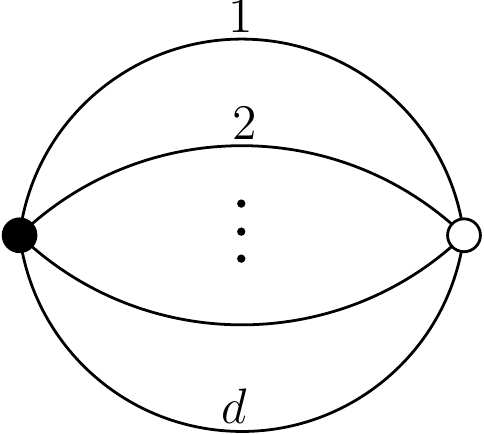}\\
(a) The 2-vertex bubble.
\end{minipage}
\quad
\begin{minipage}[b]{48mm}\centering
\includegraphics[scale=.6]{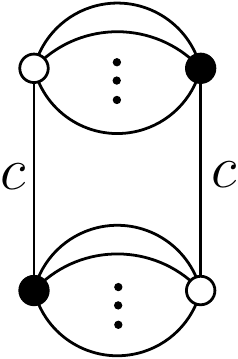}\\
(b) The 4-vertex bubble~$B_{c}$.
\end{minipage}
\quad
\begin{minipage}[b]{55mm}\centering
\includegraphics[scale=.5]{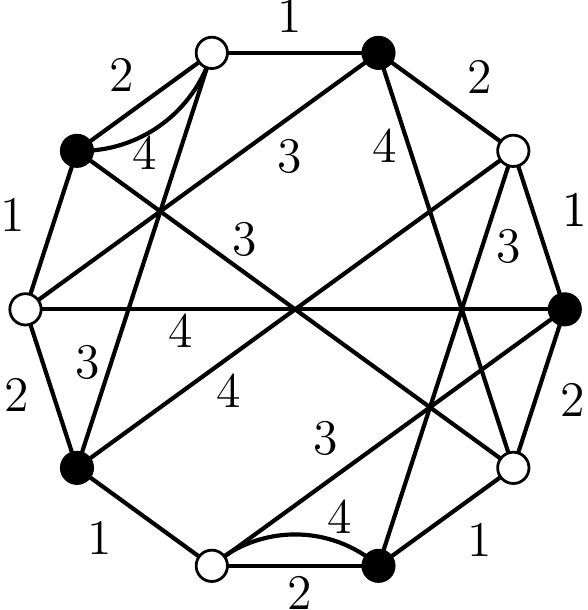}\\
(c) A 10-vertex bubble at $d=4$.
\end{minipage}
\caption{Here are some examples of bubbles. The dots indicate multiple edges.}\label{fig:Bubbles}
\end{figure}

The algebra of invariant polynomials is generated by a set of polynomials labeled by bubbles. A {\it bubble} is a connected, bipartite graph, regular of degree $d$, whose edges must be colored with a~color in $\{1, \dots, d\}$, and such that all $d$ colors are incident at each vertex (and is incident exactly once). Examples of bubbles are displayed in Fig.~\ref{fig:Bubbles}. To each bubble~$B$, a~polynomial~$P_B(T, \overline{T})$ is canonically associated, and the other way around. To do so, associate to each white (black) vertex the tensor~$T$~($\overline{T}$). If there is an edge of color $c\in\{1, \dots, d\}$ between two vertices, one identif\/ies the two indices in position $c$ of the tensors corresponding to those two vertices, and sum the index from~$1$ to~$N$, $\sum\limits_{a_c=1}^N T_{\dotsb a_c \dotsb} \overline{T}_{\dotsb a_c \dotsb}$. This way, it is easily seen that all indices are contracted between $T$s and $\overline{T}$s in a position-preserving way which
ensures unitary invariance.

There is a single quadratic invariant
\begin{gather*}
T\cdot \overline{T} = \sum_{a_1, \dots, a_d} T_{a_1 \dotsb a_d}\,\overline{T}_{a_1 \dotsb a_d},
\end{gather*}
associated with the unique bubble on two vertices (both connected by $d$ edges of all colors).

Special classes of bubbles include the melonic bubbles. To describe them, we recall that the $(d-1)$-dipole of color $c$ is the (open) graph made of two vertices connected by the $d-1$ edges of all colors except $c$, and half-edges of color $c$ incident to both vertices. A~{\it melonic bubble} on $p+2$ vertices is obtained from a melonic bubble on $p$ vertices by cutting an edge of color~$c$ into two half-edges and gluing them back to the half-edges of the $(d-1)$-dipole of color~$c$,
\begin{gather} \label{DipoleInsertion}
B = \begin{array}{@{}c@{}} \includegraphics[scale=.6]{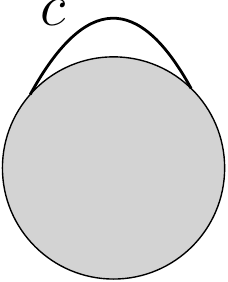} \end{array} \quad \to \quad B' = \begin{array}{@{}c@{}} \includegraphics[scale=.6]{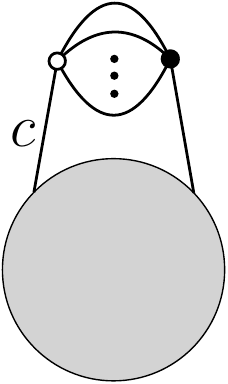} \end{array}
\end{gather}
The f\/irst melonic graphs are obtained by doing so on the bubble with two vertices. One obtains a~bubble $B_{c}$ with four vertices and $c=1, \dots, d$. Identifying the bubble with the polynomial, we can write
\begin{gather*}
P_{B_{c}}(T, \overline{T}) = \begin{array}{@{}c@{}} \includegraphics[scale=.6]{4PointBubbleColorC} \end{array}\\
\hphantom{P_{B_{c}}(T, \overline{T})}{} = \sum_{\substack{a_1, \dots, a_d \\ b_1, \dotsc, b_d}} T_{a_1 \dotsb a_{c-1} a_c a_{c+1} \dotsb a_d} \overline{T}_{a_1 \dotsb a_{c-1} b_c a_{c+1} \dotsb a_d}\, T_{b_1 \dotsb b_{c-1} b_c b_{c+1} \dotsb b_d} \overline{T}_{b_1 \dotsb b_{c-1} a_c b_{c+1} \dotsb b_d}.
\end{gather*}

Other remarkable bubbles are, for $d$ even, the so-called necklaces. They are cycles of arbitrary length where two adjacent vertices are connected by $d/2$ edges. Thus a white vertex receives the colors $\{i_1, \dotsc, i_{d/2}\}$ from a black vertex and the complementary colors from another black vertex. Denoting $M$ the $N^{d/2}\times N^{d/2}$-matrix between those two sets of colors, the necklace of length $2p$ has associated polynomial $\operatorname{Tr} (M M^\dagger)^p$ (see equation~\eqref{Necklace12}).

The counting of bubbles has been studied in \cite{BubbleCounting}. Here we are interested in counting the dif\/ferent ways to glue bubbles together instead.

\subsection[Tensor models and $(d+1)$-colored graphs]{Tensor models and $\boldsymbol{(d+1)}$-colored graphs}

Let $I$ be a f\/inite set, and $\{B_i\}_{i\in I}$ a set of bubbles. A tensor model is def\/ined by its partition function $Z$ and free energy $f$,
\begin{gather} \label{GenericZ}
Z = \exp f = \int \exp \left( -N^{d-1} T\cdot \overline{T} - \sum_{i\in I} N^{s_i} t_i P_{B_i}(T, \overline{T})\right) dT d\overline{T},
\end{gather}
and expectation of observables
\begin{gather} \label{GenericVEV}
\langle P_B(T, \overline{T}) \rangle = \frac{1}{Z} \int P_B(T, \overline{T}) \exp \left( -N^{d-1} T\cdot \overline{T} - \sum_{i\in I} N^{s_i} t_i P_{B_i}(T, \overline{T})\right) dT d\overline{T},
\end{gather}
which are all functions of the coupling constants $\{t_i\}_{i\in I}$. They also depend on the choice of the exponents $\{s_i\}_{i\in I}$. How to choose them is actually the main topic of the present article.

The relationship with edge-colored graphs comes from applying Feynman's expansion to~\eqref{GenericZ} and~\eqref{GenericVEV}. One expands the exponentials of $P_{B_i}$ (except the quadratic one) as power series and (illegally) commutes the sums with the integral. Since each~$P_{B_i}$ is a polynomial in the entries of~$T$,~$\overline{T}$, one ends up with moments of the Gaussian distribution,
\begin{gather} \label{GaussianMoment}
\int \exp \big({-}N^{d-1} T\cdot \overline{T}\big) \prod_j P_{B_{i_j}}^{b_{i_j}}(T, \overline{T}) dT d\overline{T}.
\end{gather}
They are evaluated thanks to Wick's theorem, as sums over pairings of $T$s with $\overline{T}$s. This can be represented graphically as follows. The polynomials $P_{B_{i_j}}$ appearing in the Gaussian moment are represented by their bubbles (which carry the colors from $1$ to $d$). Since $T$s are white vertices and $\overline{T}$s are black vertices, pairings between $T$s and $\overline{T}$s are then drawn as edges between black and white vertices. Those edges are assigned the color~$0$. This way, the calculation expands onto graphs, which satisfy the same def\/inition as that of bubbles, with the additional color~$0$ (and for the calculation of~$Z$, those graphs are not necessarily connected). We call them {\it $(d+1)$-colored graphs}.

The free energy therefore expands onto connected $(d+1)$-colored graphs, whose bubbles (i.e., maximally connected subgraphs with colors $1, \dotsc, d$) are chosen among the set $\{B_i\}_{i\in I}$. We call this set of (non-labeled, non-rooted) graphs ${\mathcal G}(\{B_i\}_{i\in I})$. Each graph furthermore receives a~weight. There is a free sum over a tensor index in position $c$ from $1$ to $N$ for each cycle of alternating colors $0$ and $c$. We call such a cycle a {\it face of color $0c$}. Moreover, each bubble of type $B_i$ comes with a weight $N^{s_i} t_i$, $i\in I$ and each edge of color $0$ comes with $N^{-(d-1)}$. That gives
\begin{gather} \label{FeynmanExpansionFreeEnergy}
f = \sum_{G\in {\mathcal G}(\{B_i\})} C(G) N^{\sum\limits_{c=1}^d F_{0c}(G) - (d-1) E(G) + \sum\limits_{i\in I} b_i(G) s_i} \prod_{i\in I} (-t_i)^{b_i(G)}.
\end{gather}
Here, $F_{0c}(G)$ denotes the number of faces of colors $0c$ of $G$, $E(G)$ its number of edges of color~$0$, $b_i(G)$ its number of bubbles of type~$B_i$. $C(G)$ is a numerical factor of combinatorial origin. Indeed, the Feynman expansion naturally labels the bubbles and their vertices, since all~$T$ and~$\overline{T}$ in~\eqref{GaussianMoment} are distinct. $C(G)$ is thus the number of graphs with labeled bubbles and vertices which have the same unlabeled graph $G\in {\mathcal G}(\{B_i\}_{i\in I})$, divided by $b(G)!$ (where $b(G)$ is the total number of bubbles)\footnote{In quantum f\/ield theory, normalizing the coupling constants reduces the combinatorial factor to $1/s(G)$ where $s(G)$ is the symmetry factor of $G$. However, this normalization of the coupling constants depend on their symmetry. In an interaction~$\phi(x)^n$, the $n$ copies of~$\phi(x)$ are equivalent, so the natural normalization of the coupling constant is~$1/n!$, as well known. In tensor models however this depends on the choice of bubbles.}.

Similarly, we denote ${\mathcal G}(\{B_i\};B)$ the set of connected $(d+1)$-colored graphs with a marked bubble $B$ (with labeled vertices) and all other bubbles from the set $\{B_i\}_{i\in I}$. The expectation of $P_B(T, \overline{T})$ admits a Feynman expansion onto the graphs of ${\mathcal G}(\{B_i\};B)$,
\begin{gather*}
\langle P_B(T, \overline{T}) \rangle = \sum_{G\in {\mathcal G}(\{B_i\};B)} C(G) N^{\sum\limits_{c=1}^d F_{0c}(G) - (d-1) E(G) + \sum\limits_{i\in I} b_i(G) s_i} \prod_{i\in I} (-t_i)^{b_i(G)}.
\end{gather*}
In that case, the bubble $B$ is not counted among the $b_i(G)$ (it may be dif\/ferent from all~$B_i$ for $i\in I$ anyway). Moreover, the combinatorial factor~$C(G)$ is now evaluated with labeled vertices on~$B$.

It will be convenient to def\/ine the {\it power of $G\in {\mathcal G}(\{B_i\}_{i\in I})$} as
\begin{gather*}
\delta_{\{s_i\}_{i\in I}}(G) = F(G) - (d-1) E(G) + \sum_{i\in I} b_i(G) s_i, \qquad \text{with} \quad F(G) = \sum_{c=1}^d F_{0c}(G)
\end{gather*}
being the total number of faces, so that a graph $G$ in the above expansions like \eqref{FeynmanExpansionFreeEnergy} is counted with a weight $N^{\delta_{\{s_i\}_{i\in I}}(G)} \prod\limits_{i\in I} (-t_i)^{b_i(G)}$.
\begin{itemize}\itemsep=0pt
\item A tensor model is said to have a $1/N$-expansion if
\begin{gather*}
f \leq A N^D,
\end{gather*}
for some $N$-independent quantities $A$, $D$, i.e., the exponent of $N$ in the summand in~\eqref{FeynmanExpansionFreeEnergy} is bounded,
\begin{gather*}
\exists \, D \quad \forall\, G\in{\mathcal G}(\{B_i\}) \quad \delta_{\{s_i\}_{i\in I}}(G) \leq D.
\end{gather*}
\item It is furthermore said that the large $N$ limit is non-trivial if there is an inf\/inite family of graphs which contribute to the limit $\lim\limits_{N\to \infty} f/N^D$.
\end{itemize}

Those two conditions (existence of a $1/N$-expansion and non-triviality of the large $N$ limit) will be used to determine the values of the exponents $\{s_i\}_{i\in I}$. Indeed, if some of the $s_i$ are too large, the exponent of~$N$ in~\eqref{FeynmanExpansionFreeEnergy} can get larger and larger as the number of bubbles grows. Requiring the existence of a $1/N$-expansion enforces $s_i$ not to be too large. The reason the notion of non-trivial large $N$ limit is introduced is that by taking the exponents~$s_i$ suf\/f\/iciently small, one can easily build tensor models which do have a $1/N$-expansion. However, the large $N$ limits obtained this way are typically uninteresting: only a f\/inite number of graphs contributes to each order of the $1/N$-expansion. Then $f/N^D$ is a polynomial in the couplings $\{t_i\}$. We are interested on the contrary in the cases where $f$ develops singularities. This is possible only when $\lim\limits_{N\to \infty} f/N^D$ is an inf\/inite sum of graphs.

\begin{Remark} The coef\/f\/icient $C(G)$ is very important in the explicit evaluation of the free energy. However, we will not have to care too much about it in this article. Indeed, in most sections we will be interested in the existence of $1/N$-expansions and in f\/inding the families of graphs which contribute to $\lim\limits_{N\to \infty} f/N^D$. The coef\/f\/icient $C(G)$ is then irrelevant. As for the sections dealing with enumeration, we will use Schwinger--Dyson-like techniques which bypass the explicit analysis of~$C(G)$. Moreover, our most developed section on enumeration is Section~\ref{sec:Quartic} devoted to the quartic case. In that case, $C(G) = 2^{b(G)}$, which can be re-absorbed by a redef\/inition of the couplings $t_i \to t_i/2$ (this is because quartic bubbles have two equivalent black, or white, vertices).
\end{Remark}

\section{Gurau's degree theorem} \label{sec:Degree}

Tensor models were revived thanks to Gurau's discovery that there is a value of $s_i$ which provides tensor models with a $1/N$-expansion. This was proved in \cite{LargeN1, LargeN3, LargeN2} by a detailed analysis of all $(d+1)$-colored graphs. It was then adapted to the colored graphs~${\mathcal G}(\{B_i\})$ with prescribed bubbles in~\cite{Uncoloring}. We state the theorem is this context.

\begin{Theorem} \label{thm:Degree}
For any finite set of bubbles $\{B_i\}_{i\in I}$ and any graph $G\in {\mathcal G}(\{B_i\})$,
\begin{gather} \label{GurauDegreeThm}
\delta_{\{s_i = d-1\}}(G) = \sum_{c=1}^d F_{0c}(G) - (d-1) \bigg(E(G) - \sum_{i\in I} b_i(G) \bigg) \leq d.
\end{gather}
\end{Theorem}

Equivalently, one def\/ines {\it Gurau's degree}
\begin{gather*}
\omega(G) \equiv d - \sum_{c=1}^d F_{0c}(G) + (d-1) \bigg(E(G) - \sum_{i\in I} b_i(G) \bigg),
\end{gather*}
and the theorem states that $\omega(G)\geq0$.

The key application of the theorem is to notice that with the choice $s_i = d-1$ for all $i\in I$, each graph $G$ in the summand of~\eqref{FeynmanExpansionFreeEnergy} is weighted like $N^{d - \omega(G)}$. In particular, the free energy is then bounded. In other words, choices
\begin{gather*}
s_i \leq d-1
\end{gather*}
ensure the existence of the $1/N$-expansion for any bubble. The bound obtained on the free energy is $f \sim N^d$. That is natural since $N^d$ is the total number of degrees of freedom of the random tensor, and the free energy is extensive.

The question is then to f\/ind whether the large $N$ limit obtained for $s_i = d-1$ is non-trivial. Before that, we show that the above theorem encompasses the well-known 2D case, where the degree of a graph reduces to (twice) the genus of a map.

\subsection[The case $d=2$: combinatorial maps]{The case $\boldsymbol{d=2}$: combinatorial maps}

Bubbles at $d=2$ are connected 2-colored graphs, hence they are cycles, characterized by their number of vertices. A graph $G\in {\mathcal G}(\{B_p\}_{p\in 2\mathbb{N}})$, where $B_p$ is the cycle with $2p$ vertices, consists in cycles connected by edges of color $0$.

First, one can represent $G$ as a bipartite map $M_{\text{col}}(G)$ with colored edges. Indeed, draw~$G$ such that the cyclic, counter-clockwise order of the colors around each black vertex is $(0, 1, 2)$, and $(0, 2, 1)$ around each white vertex. The faces of the map obtained this way are by construction partitioned into three sets of colored faces: those with alternating colors~$01$, colors~$02$ and colors~$12$. They are obviously in bijection with the corresponding faces of~$G$.

Notice that the bubbles in $G$ form a disjoint union of cycles which account for all edges of colors $1$ and $2$. Those cycles are the faces of colors~$12$ in $M_{\text{col}}(G)$. They can be shrunk to a~point, while keeping track of the cyclic order of the edges of color $0$ incident to the face. This way, one obtains for each face of colors~$12$ and degree~$2p$ a locally-embedded vertex of degree~$2p$. The remaining edges are the edges of color~$0$, whose color can now be erased. One gets a combinatorial (non-colored) map~$M(G)$ whose vertices have the degrees of the original bubbles.

Topologically, $M(G)$ is obtained from $M_{\text{col}}(G)$ through local homotopy transformations, so that they both have the same genus~$g(M(G))$.

This mapping can be inverted to associate a $3$-colored graph to each map whose vertices have even degrees, up to the symmetries which exchange black with white vertices and edges of colors~1 and~2.

Each graph $G$ comes with an exponent of $N$, given in equation~\eqref{GurauDegreeThm}. It can be rewritten in terms of the combinatorial quantities of~$M(G)$,
\begin{gather*}
F_{01}(G) + F_{02}(G) - E(G) + \sum_{p\geq 2} b_p(G)\\
\qquad{} = F(M(G)) - E(M(G)) + V(M(G)) = 2 - 2 g(M(G)).
\end{gather*}
Indeed, the faces of $M(G)$ are the faces of colors $01$ and $02$ of $G$, its edges are the edges of color~$0$ of~$G$ and its vertices are the bubbles of~$G$. Therefore, each graph in the Feynman expansion is weighted by $N^{2 - 2g(M(G))}$ where~$g(M(G))$ is the genus of the map $M(G)$. Gurau's degree theorem thus reproduces at $d=2$ the famous topological expansion of matrix models. This means that it is a genuine generalization of the~$2D$ case.

\subsection[Large $N$ limit for $d\geq 3$]{Large $\boldsymbol{N}$ limit for $\boldsymbol{d\geq 3}$}

\subsubsection{Melonicity}

At $d=2$, the large $N$ limit thus consists in planar maps. However, the situation becomes drastically dif\/ferent for $d\geq 3$. In Section~\ref{sec:Bubbles} we have def\/ined melonic bubbles. {\it Melonic graphs} are constructed exactly the same way with one additional color. Notice that the bubbles of a~melonic graph are melonic bubbles.

\begin{Theorem} \label{thm:Melons}
Let $G\in {\mathcal G}(\{B_i\}_{i\in I})$. Then
\begin{gather*}
\sum_{c=1}^d F_{0c}(G) - (d-1) \bigg(E(G) - \sum_{i\in I} b_i(G) \bigg) = d
\end{gather*}
$($which is the vanishing of Gurau's degree$)$ if and only if $G$ is a melonic graph. This forces all bubbles appearing in~$G$ to be melonic too.
\end{Theorem}

The proof is given in \cite{Uncoloring}. This theorem shows that the large $N$ limit obtained with $s_i = d-1$ for all bubbles $B_i$ in the action is non-trivial only if some of those bubbles are melonic.

\subsubsection{Universality}

Tensor models can be solved at large $N$ just like matrix models, using either direct combinatorial arguments \cite{Universality}, or using the Schwinger--Dyson equations \cite{SDE} (more details on the Schwinger--Dyson equations of tensor models can be found in \cite{TreeAlgebra, BubbleAlgebra}). One f\/irst proves, using either methods, the following universality theorem, which f\/irst appeared in \cite{Universality}.

\begin{Theorem} \label{thm:Universality}
Let $B'$ be a bubble with a $(d-1)$-dipole and $B$ the same bubble after the dipole removal, like in equation~\eqref{DipoleInsertion}. Then,{\samepage
\begin{gather} \label{DipoleFactorization}
\langle P_{B'}(T,\overline{T}) \rangle = G_2(\{t_i\}_{i\in I}) \langle P_{B}(T, \overline{T}) \rangle, \qquad \text{where} \quad
G_2(\{t_i\}_{i\in I}) = \frac1N \langle T\cdot \overline{T} \rangle
\end{gather}
is the large $N$, $2$-point function.}

As a consequence, if $B$ is melonic $($meaning built out of recursive $(d-1)$-dipole insertions$)$, then
\begin{gather} \label{MelonicGaussianExpectation}
\frac1N \langle P_B(T, \overline{T}) \rangle = G_2(\{t_i\}_{i\in I})^{p(B)},
\end{gather}
where $p(B)$ is the number of black vertices of~$B$.
\end{Theorem}

In the theorem and in the remaining of the section, all equalities hold in the large $N$ limit. Equation~\eqref{DipoleFactorization} can be explained as follows. Denote $v$, $\bar{v}$ the white and black vertices of the $(d-1)$-dipole in $B'$. All graphs $G\in {\mathcal G}(\{B_i\}_{i\in I};B')$ which contribute to the expectation of~$P_{B'}$ at large~$N$ must contain a 2-point function between~$v$ and~$\bar{v}$, i.e., a contribution such that cutting the two edges of color 0 incident to~$v$ and $\bar{v}$ disconnects~$G$. The set of all contributions surviving the large $N$ limit therefore factorizes: all the 2-point functions connecting $v$ to $\bar{v}$ which sum up to $G_2(\{t_i\}_{i\in I})$, and all the large $N$ contributions which do not see the dipole, i.e., the expectation of~$P_B$. Graphically,
\begin{gather*}
\langle P_{B'}(T, \overline{T}) \rangle = \left\langle \begin{array}{@{}c@{}}\includegraphics[scale=.5]{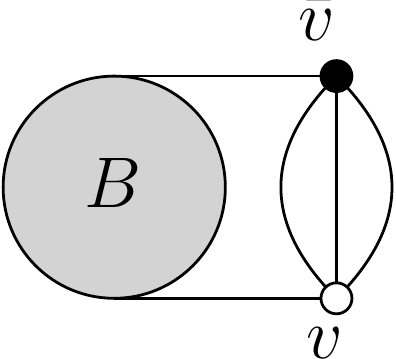} \end{array} \right\rangle = \begin{array}{@{}c@{}} \includegraphics[scale=.5]{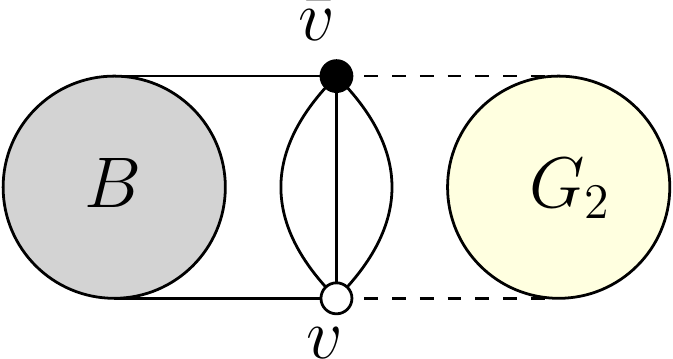} \end{array} = G_2(\{t_i\}_{i\in I}) \langle P_B(T, \overline{T}) \rangle.
\end{gather*}

In fact the theorem in~\cite{Universality} shows that tensor models are \emph{Gaussian} in the large $N$ limit, with covariance given by $G_2$, i.e., all expectations factorize as products of the large~$N$, 2-point function. In a Gaussian distribution, the expectation of a polynomial reads as a sum over pairings of~$T$s with $\overline{T}$s. For a melonic bubble, there is a single pairing which survives the large~$N$ limit. It can be found iteratively by pairing the vertices of each $(d-1)$-dipole, then removing those dipoles and repeating the process.

Note that \eqref{MelonicGaussianExpectation} shows that the expectation of a melonic bubble only depends on its number of vertices, and not on its particular structure.

\subsubsection{Enumeration and continuum limit}

If $B$ is melonic, it has a $(d-1)$-dipole with vertices $v$, $\bar{v}$. We denote $B/ (v,\bar{v})$ the melonic bubble obtained by replacing the dipole with an edge. We also def\/ine the gluings of bubbles: remove the vertex~$v$ of~$B$, which leaves half-edges of color $1, \dots, d$ hanging out from black vertices, and similarly remove a black vertex~$\bar{v}_i$ from a bubble~$B_i$. One can connect the half-edges which have the same colors, so as to get a new (connected) bubble, denoted $B\cdot_{(v, \bar{v}_i)} B_i$. The gluing of two melonic bubbles still is melonic. The Schwinger--Dyson equations read~\cite{SDE}
\begin{gather*}
\langle P_{B/ (v,\bar{v})}(T, \overline{T})\rangle - \langle P_B(T, \overline{T}) \rangle - \sum_{i\in I} t_i \sum_{\bar{v}_i \in B_i} \langle P_{B\cdot_{(v, \bar{v}_i)} B_i}(T, \overline{T}) \rangle = 0.
\end{gather*}
This {\it a priori} complicated set of equations simplify drastically thanks to the universality theo\-rem~\eqref{MelonicGaussianExpectation}, as all expectations then only depend on the numbers of vertices. They all collapse onto an equation which determines the 2-point function,{\samepage
\begin{gather} \label{MelonicG2}
1 - G_2(\{t_i\}_{i\in I}) - \sum_{i\in I} p_i t_i G_2(\{t_i\}_{i\in I})^{p_i} = 0,
\end{gather}
where $p_i$ is the number of black vertices of $B_i$.}

This is a polynomial equation on $G_2$. Let us rescale all the couplings $\{t_i\}_{i\in I}$ with $\lambda$\footnote{It is also convenient and customary to instead rescale the action by a global parameter $1/\lambda$, which turns the~$1$ in~\eqref{MelonicG2} into~$\lambda$ so that the equation reads $\lambda = F(G_2)$ for some polynomial $F$. Such an equation is particularly easy to analyze.}. Thus (dropping the explicit dependence on the couplings except $\lambda$),
\begin{gather*}
G_2(\lambda) = \sum_{\substack{G \in {\mathcal G}(\{B_i\}_{i\in I}\\ \text{melonic with a marked edge}}} C(G) \lambda^{b(G)} \prod_{i\in I} (-t_i)^{b_i(G)},
\end{gather*}
i.e., $\lambda$ counts the melonic graphs $G$ with a marked edge of color 0, with respect to the number of bubbles $b(G)$. One obtains
\begin{gather} \label{MelonicCounting}
1 - G_2(\lambda) - \lambda \sum_{i\in I} p_i t_i G_2(\lambda)^{p_i} = 0.
\end{gather}
A standard theorem on algebraic generating functions \cite{AC} shows that for generic couplings, $G_2$~has a dominant singularity of the form
\begin{gather*}
G_2(\lambda) \sim \sqrt{\lambda_c - \lambda},
\end{gather*}
where $\lambda_c$ is the radius of convergence of $G_2$. The free energy $f(\lambda)$ therefore behaves as
\begin{gather*}
f(\lambda) \sim (\lambda_c - \lambda)^{2 - \gamma}, \qquad \text{for} \quad \gamma = \frac12.
\end{gather*}
The critical exponent $\gamma$ is known as the {\it entropy} exponent. The value~$1/2$ is quite universal for algebraic functions, and typical of trees. $\gamma=1/2$ is in fact known as the branched polymer exponent. It was later proved that the geometry of the melonic graphs also converges to that of the continuous random tree~\cite{BP}.

By tuning the couplings $\{t_i\}_{i\in I}$ to specif\/ic values, one can in addition reach the multi-critical exponents of branched polymers, $\gamma = 1 - 1/m$, for~$m$ integer and $m\geq 2$,~\cite{Uncoloring}.

\section[New large $N$ limits]{New large $\boldsymbol{N}$ limits} \label{sec:New1/N}

For tensor models with melonic bubbles and $s = d-1$,
\begin{itemize}\itemsep=0pt
\item in combinatorial terms: melonic graphs (equivalent to trees) dominate the Feynman expansions of all expectations,
\item in probabilistic terms: the large $N$ limit is a Gaussian distribution with covariance~$G_2$.
\end{itemize}
Those are two related facts, and it is natural to expect that escaping the branched polymer behavior of melonic graphs comes with a non-Gaussian large $N$ limit.

This clearly has to be done using {\it non-melonic} bubbles. Theorems~\ref{thm:Degree} and~\ref{thm:Melons} however fall short of a description of what happens with non-melonic bubbles. Indeed, one does not know the minimal value of the degrees of the graphs built from arbitrary non-melonic bubbles. The degree certainly increases with the number of bubbles in the graphs. If it does so linearly, one could then f\/ind a value of $s_i>d-1$ which would lead to a~non-trivial large~$N$ limit.

\subsection{An example} \label{sec:NonGaussianExample}

Let us f\/irst show that the above scenario, that of a non-trivial, non-Gaussian large $N$ limit, is valid on an example. It relies on the fact that ordinary matrix models have non-branched polymer, non-Gaussian, large $N$ limits. Moreover, they can be turned to tensor models, by doubling the indices into pairs of indices for instance, $M_{AB} = T_{a_1 a_2 a_3 a_4}$, with $A = (a_1, a_3)$, $B = (a_2, a_4)$. Through this correspondence one has
\begin{gather} \label{Necklace12}
\operatorname{Tr}(M M^\dagger)^p = P_{B_p}(T, \overline{T}) = \begin{array}{@{}c@{}} \includegraphics[scale=.65]{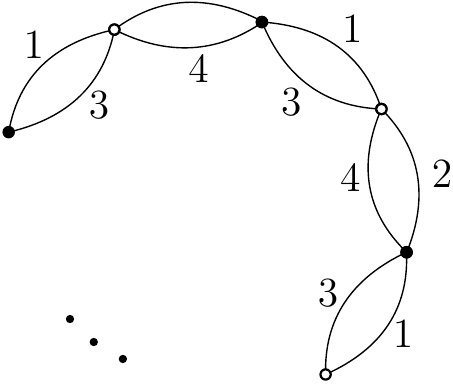} \end{array}
\end{gather}
where we represent the polynomial by its bubble $B_p$ with $2p$ vertices, and thus
\begin{gather*}
\int dMdM^\dagger \, e^{-N^2 \sum_{p\geq 1} t_p \operatorname{Tr}(MM^\dagger)^p} = \int dT d\overline{T}\, e^{-N^2\Big(t_1 T\cdot \overline{T} + \sum\limits_{p\geq 2} t_p P_{B_p}(T, \overline{T})\Big)}.
\end{gather*}
Rescaling $T$, $\overline{T}$ by $N^{1/2}$ so that the factor in front of $T\cdot \overline{T}$ becomes $N^{d-1}$ at $d=4$, as in~\eqref{GenericZ} gives
\begin{gather*}
\int dT d\overline{T}\, \exp \bigg({}-N^3 t_1 T\cdot \overline{T} + \sum_{p\geq 2} N^{2+p} t_p P_{B_p}(T, \overline{T}) \bigg).
\end{gather*}
In other words, we have found some exponents
\begin{gather} \label{NecklaceEnhancement}
s_p = p+2 > d-1,
\end{gather}
for the bubbles $B_p$, $p\geq 2$, such that $i)$ there is a $1/N$-expansion, $ii)$ the large $N$ limit is non-Gaussian (the set of ``planar'' graphs, in the appropriate sense, as it can be solved as an ordinary matrix model~\cite{matrix}). Furthermore, the trick we have just described can be applied to any tensor with an even number of indices (the matrix $M_{AB}$ would have for instance an index $A = (a_1, a_3, a_5, \dotsc)$ containing all the odd colors while $B = (a_2, a_4, a_6, \dots)$ contains all the even colors).

If the exponent $s = d-1$ of Theorem~\ref{thm:Degree} had been used in front of the bubbles $B_p$ shown in~\eqref{Necklace12} in the action, the large~$N$ limit would have been trivial (a Gaussian with covariance~$1/t_1$) because they are not melonic. We can in fact check that the degree of the graphs grows with the number of bubbles. From its def\/inition, Gurau's degree is
\begin{gather*}
\omega(G) = 4 - \sum_{c=1}^4 F_{0c}(G) + 3 \sum_{p \geq 2} (p - 1) b_p(G),
\end{gather*}
where $b_p(G)$ is the number of bubbles of type $B_p$ in $G$, and we have used $E(G) = \sum\limits_{p \geq 2} p b_p(G)$. Notice that any graph $G$ can be obtained from a 3-colored graph with colors $0$, $1$, $2$ by doubling the edges of colors~1 and~2. The corresponding 3-colored graph represents a surface of genus~$g(G)$ with $2-2g(G) = F_{01}(G) + F_{02}(G) - \sum\limits_{p\geq 2} (p-1) b_p(G)$. Moreover, by construction, $F_{03}(G) + F_{04}(G) = F_{01}(G) + F_{02}(G)$, and then
\begin{gather*}
\omega(G) = 4 g(G) + \sum_{p\geq 2} (p-1) b_p(G) \geq \sum_{p\geq 2} (p-1) b_p(G).
\end{gather*}
It comes that among the set of planar graphs, the degree indeed increases linearly with the number of bubbles.

Choosing $s_p = p+2 > d-1$ to be bubble-dependent and larger than $d-1$ enhances the bubbles~$B_p$ so that each graph $G$ receives a weight $N^{4 - 4g(G)}$ instead of $N^{4-\omega(G)}$. The large $N$ limit consists of the set of graphs for which $g(G)=0$, which can be made of an arbitrary number of bubbles. Therefore the couplings $t_p$ contribute to a non-trivial large~$N$ limit, while preserving the~$1/N$-expansion.

\subsection{The key question: do enhancements exist?}

The key question we will henceforth discuss in this article is the following. Given a non-melonic bubble $B$, is there a value of $s_B > d-1$ such that
\begin{gather*}
f_B(t_B) = \ln \int dT d\overline{T}\, \exp \bigl({-}N^{d-1} T\cdot \overline{T} - N^{s_B} t_B P_B(T, \overline{T}) \bigr)
\end{gather*}
admits a $1/N$-expansion? If so, we say that $s_B$ is the {\it enhancement} of~$B$. When it exists, one can then ask what the large $N$ limit is? Also, if two bubbles have enhancements, can they be used together in the action like in \eqref{GenericZ}?

The above free energy has the Feynman expansion
\begin{gather*}
f_B(t_B) = \sum_{G\in {\mathcal G}(B)} C(G) N^{\sum\limits_{c=1}^d F_{0c}(G) - (d-1)E(G) + s_B b(G)} (-t_B)^{b(G)},
\end{gather*}
where ${\mathcal G}(B)$ denotes the set of connected $(d+1)$-colored graphs whose bubbles are copies of $B$, and $b(G)$ is the number of such bubbles in $G$. Assume that $s_B$ is f\/ixed so that $f_B$ is bounded by some f\/ixed power of $N$ and also has a non-trivial large $N$ limit, i.e., there exists an inf\/inite family of graphs ${\mathcal G}_{\max}(B)$ maximizing the power of $N$. Then, the enhancement is {\it unique}. Indeed, if $\epsilon>0$ is added to $s_B$, then the graphs in ${\mathcal G}_{\max}(B)$ will receive a weight which grows like $N^{\epsilon b(G)}$, i.e., which diverges with the number of bubbles. This would ruin the $1/N$-expansion. Therefore, there is a unique maximal enhancement.

Observe that $E(G) = p(B) b(G)$, where $p(B)$ denotes the number of black vertices of $B$. Thus, the number of edges of color $0$ is determined by and scales linearly with the number of bubbles. The power of a graph rewrites
\begin{gather} \label{Power}
\delta_{s_B}(G) = F(G) - \bigl[(d-1) p(B) - s_B \bigr] b(G).
\end{gather}
As a consequence, establishing the existence of a $1/N$-expansion of~$f_B$, which means f\/inding a~bound on~$f_B$, hence on $\delta_{s_B}(G)$, is equivalent to {\it determining how the number of faces grows with the number of bubbles for those graphs which maximize that number of faces at fixed number of bubbles}.

Therefore if for a given $B$ one can identify combinatorially the graphs which maximize the number of faces at f\/ixed number of bubbles, then one can f\/ind the maximal value of the enhancement $s_B$.

\section{Bubble gluing and inherited enhancements} \label{sec:BubbleGluing}

Before we present some strategies to f\/ind the graphs which maximize the number of faces, let us assume that we have done so for a~given bubble $B$, and let us try and use this knowledge to identify the enhancements for other bubbles.

More precisely, we assume that there exists $s_B$ such that $f_B$ admits a $1/N$-expansion with a~non-trivial large $N$ limit. This means that $\delta_{s_B}(G)$ in~\eqref{Power} is bounded for all $G$ by $\delta_{\max}(B) = \max\limits_{G'\in {\mathcal G}(B)} \delta_{s_B}(G')$ and there is an inf\/inite family of graphs ${\mathcal G}_{\max}(B) \subset {\mathcal G}(B)$ whose graphs reach the bound on $\delta_{s_B}(G)$,
\begin{gather*}
{\mathcal G}_{\max}(B) = \{ G\in {\mathcal G}(B),\,\delta_{s_B}(G) = \delta_{\max}(B)\}.
\end{gather*}

Consider colored graphs with {\it free vertices}, i.e., vertices which do not have an incident edge of color $0$ (but they do have all the other colors $1, \dotsc, d$). A~bubble can thus be seen as a~colored graph with only free vertices. Gluing two bubbles with respectively $2p$ and $2p'$ vertices via a~single edge of color~$0$ leads to a graph with $2(p+p'-1)$ free vertices, and so on.

We denote ${\mathcal G}_k(B)$ the set of graphs whose bubbles are all copies of $B$ and with $2k$ free vertices. Clearly, all graphs in ${\mathcal G}_k(B)$ can be obtained (non-uniquely) by removing edges of color $0$ from graphs in~${\mathcal G}(B)$.

Recall that a face of colors $(0c)$ is a cycle alternating the colors $0$ and~$c$. If $H\in {\mathcal G}_k(B)$, we denote its number of faces $F(H)$. Since it has free vertices, there are also paths alternating the colors $0$ and $c$ which are not closed.

\begin{figure}[t]
\centering\includegraphics[scale=.6]{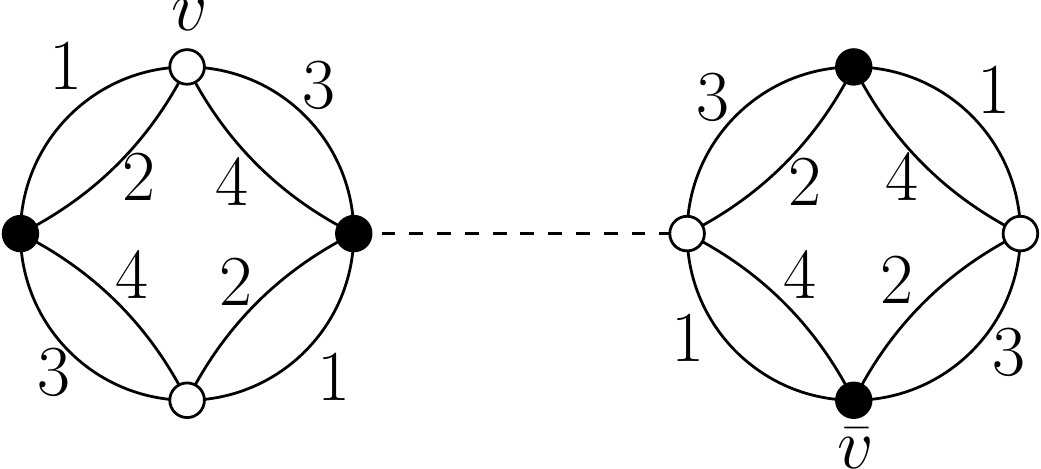} \qquad\qquad \includegraphics[scale=.6]{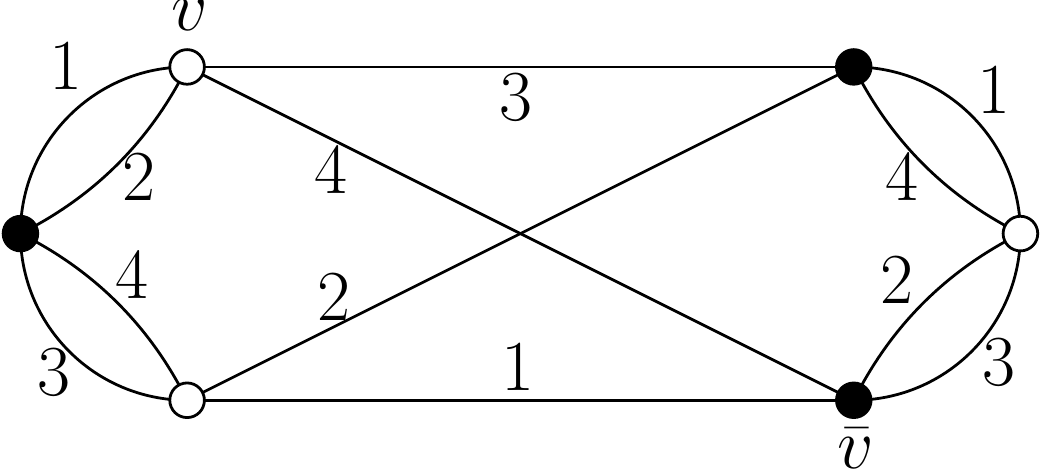}
\caption{An example of a graph $H$ with free vertices on the left and its boundary bubble~$\partial H$ on the right. The vertices of $\partial H$ are the free vertices of $H$. The edges of color $c$ in $\partial H$ are the paths of colors~$(0c)$ in~$H$. All edges between free vertices in $H$ remain as such in $\partial H$. For instance, there is path with colors~$4$,~$0$,~$4$ from the top left white vertex~$v$ to the right bottom black vertex $\bar{v}$ of $H$. It gives rise to an edge of color~$4$ between those vertices in~$\partial H$.}\label{fig:BoundaryBubble}
\end{figure}

Any graph $H\in {\mathcal G}_k(B)$ def\/ines naturally a bubble with $2k$ vertices which we denote $\partial H$ and call the {\it boundary bubble of~$H$}. It is obtained in the following way. The vertices of $\partial H$ are the free vertices of~$H$ (keeping the black and white coloring). There is an edge of color~$c$ in $\partial H$ bet\-ween~$v$ and~$\bar{v}$ if there is a path alternating the colors~$0$ and~$c$ bet\-ween~$v$ and~$\bar{v}$ in~$H$. The operation of taking the boundary bubble of two bubbles glued via an edge of color $0$ is one of the two operations necessary to formulate the Schwinger--Dyson equations of tensor models (analogous to the loop equations of matrix models, or Tutte's equations of combinatorial maps)~\cite{BubbleAlgebra}. An example is shown in Fig.~\ref{fig:BoundaryBubble}.

Here, our interest in the boundary bubble is the following. Assume that $H\in{\mathcal G}_k(B)$ is a~subgraph of $G\in{\mathcal G}(B)$. $G$ has two types of faces: those restricted to $H$ (there are $F(H)$ of them) and those which have at least one edge not in~$H$ ($F(G,H) = F(G) - F(H)$ of them). Let us form a new colored graph $G'$ with no free vertices by replacing $H\subset G$ with $\partial H$. The faces which were conf\/ined to~$H$ have disappeared in $G'$. However, the faces which went through a~path in $H$ now follow the edges of $\partial H$ with the same entry and exit vertices, by def\/inition of~$\partial H$. For instance, if $H$ is the graph on the left of Fig.~\ref{fig:BoundaryBubble}, then there is a face of colors~$04$ in~$G$ which enters~$H$ via the top left white vertex~$v$ and leaves through the right bottom black vertex~$\bar{v}$. In~$G'$ this face now goes straight between the two vertices via the edge of color $4$ between~$v$ and~$\bar{v}$ in~$\partial H$ on the right of Fig.~\ref{fig:BoundaryBubble}. Finally, the faces which did not touch~$H$ are unchanged in~$G'$. This shows that
\begin{gather*}
F(G') = F(G,H).
\end{gather*}
In other words, the boundary graph $\partial H$ keeps the structure seen by the rest of $G$ while forgetting about the internal structure of $H$ (the non-free vertices).

Consider now $H\in {\mathcal G}_{p(H)}(B)$ with $b(H)$ copies of $B$, and the set ${\mathcal G}(H)\subset {\mathcal G}(B)$ obtained by gluing copies of $H$ via edges of color $0$. Denote further ${\mathcal G}_{\max}(H) \subset {\mathcal G}_{\max}(B)$ those which reach the maximal value of the power $\delta_{s_B}$ and assume that it is an inf\/inite set. It implies that we have found the graphs which maximize the number of faces at f\/ixed number of bubbles for the bubble~$\partial H$ as well as its enhancement $s_{\partial H}$, as shown below.

First, the set ${\mathcal G}(\partial H)$ can be seen as a subset of ${\mathcal G}(B)$. Let $G\in {\mathcal G}(\partial H)$, then there exists $\tilde{G}$ in ${\mathcal G}(H) \subset {\mathcal G}(B)$ such that replacing all copies of $H$ in $\tilde{G}$ with $\partial H$, one obtains $G$. We denote~$b(G)$ the number of copies of $\partial H$ in $G$. Then
\begin{gather*}
F(G) = F(\tilde{G}) - b(G) F(H).
\end{gather*}
We know that $F(\tilde{G})$ is bounded like
\begin{gather*}
F(\tilde{G}) \leq \bigl[(d-1)p(B) - s_B\bigr] b(\tilde{G}) + \delta_{\max}(B)
\end{gather*}
and moreover the number of copies of $B$ in $\tilde{G}$ is $b(\tilde{G}) = b(G)b(H)$, hence
\begin{gather*}
F(G) \leq \bigl(\bigl[(d-1)p(B) - s_B\bigr] b(H) - F(H)\bigr) b(G) + \delta_{\max}(B).
\end{gather*}
We thus f\/ind that the power
\begin{gather*}
\delta_{s_{\partial H}}(G) = F(G) - \bigl[(d-1)p(\partial H) - s_{\partial H}\bigr] b(G)
\end{gather*}
def\/ined with
\begin{gather} \label{InheritedEnhancement}
s_{\partial H} = (d-1)\bigl(p(\partial H) - p(B) b(H)\bigr) + s_B b(H) + F(H),
\end{gather}
is bounded. There is moreover an inf\/inite family ${\mathcal G}_{\max}(\partial H)$ of graphs which reach the maximal value of the power, obtained by replacing $H$ with $\partial H$ in ${\mathcal G}_{\max}(H)$.

Notice that the above value of $s_{\partial H}$ is \emph{inherited} from~$B$, i.e., def\/ined in a way which depends on the representative $H$ built from $B$, and not only on the bubble $\partial H$ itself. However, uniqueness of the enhancement guarantees that we would f\/ind the same value of $s_{\partial H}$ for any other way of representing this bubble as a boundary bubble.

Of course, the above reasoning applies with several bubbles $\{B_i\}_{i\in I}$ instead of the single bubble $B$. Similarly, one can then add $\partial H$ with its enhancement to the action including the bubbles $\{B_i\}_{i\in I}$ and still have a non-trivial large $N$ limit.

Since the technique we have presented here requires to generate $\partial H$ as a boundary bubble from another bubble $B$, one might ask what bubbles can be generated that way. The answer is all of them: the quartic model with its $d$ possible quartic melonic bubbles generate all bubbles as boundary bubbles, as shown in \cite{StuffedWalshMaps}. However, only the melonic bubbles are boundary graphs of subgraphs in ${\mathcal G}_{\max}(\{B_i\})$~-- all other bubbles are generated at higher orders of the $1/N$-expansion (and their enhancement cannot be found from the quartic model). Still, let us check our formula~\eqref{InheritedEnhancement} for the inherited enhancement on the melonic bubbles (for which we know that $s = d-1$).

It is easy to see that a melonic bubble $B$ with $2p$ vertices can be obtained as the boundary graph of a gluing of quartic melonic bubbles in a tree-like fashion, meaning $F(H) = 0$ and $b(H) = p-1$. Then formula~\eqref{InheritedEnhancement} gives $s = (d-1)(p - 2b(H)) + (d-1)b(H) = d-1$, as expected.

Let us also derive the enhancements of the examples in Section~\ref{sec:NonGaussianExample} from the quartic case. Since those are not melonic bubbles, we use the quartic ``necklace'' bubbles,
\begin{gather*}
\partial \left(\begin{array}{@{}c@{}} \includegraphics[scale=.75]{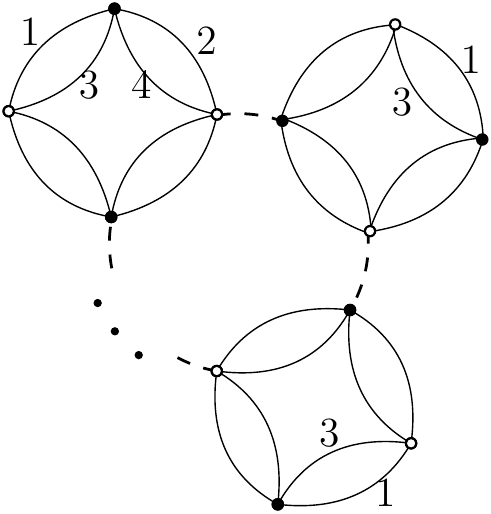} \end{array}\right) = \begin{array}{@{}c@{}}\includegraphics[scale=.75]{Necklace2pVertices}\end{array}
\end{gather*}
Formula \eqref{InheritedEnhancement} is applied with $p(\partial H) = b(H) = p$, $F(H)=2$ (one face of colors $(03)$ and one face of colors $(01)$) and $s_B = 4$ (the enhancement of the quartic necklace), and gives $s_p = -3p + 4p + 2 = p+2$, as already found from the matrix model side in equation~\eqref{NecklaceEnhancement}.

It becomes more interesting when combining for instance melonic and necklace quartic bubbles. The graphs which maximize the number of faces at f\/ixed number of bubbles were found in~\cite{MelonoPlanar}. This then allowed the authors to f\/ind the enhancements and the graphs in ${\mathcal G}_{\max}$ for an inf\/inite family of bubbles which mixes melonic and necklace features and called ``trees of necklaces''. The enumeration of the graphs in ${\mathcal G}_{\max}$ (i.e., the calculation of the free energy) was moreover performed in~\cite{MelonoPlanar} (it leads to the same universality classes as those presented in Section~\ref{sec:Quartic}).

\section{Enhancing bubbles using color slices} \label{sec:Slices}

There are some cases where the graphs which maximize the number of faces at f\/ixed number of bubbles are easily found. Let us start with an example. Suppose we have a non-melonic bubble~$B$ at $d=6$, with~$p(B)$ black vertices, whose subgraph with colors $1$, $2$, $3$ is melonic and whose subgraph with colors $4$, $5$, $6$ is melonic too. Then, if $s_B=d-1$, the tensor model with that bubble as interaction is independent of the bubble coupling at large $N$ because the degree is always positive, $\sum\limits_{c=1}^6 F_{0c}(G) - 5(p(B)-1) b(G) = 6 - \omega(G) <6$. However, if~$B$ is such that~${\mathcal G}(B)$ contains $7$-colored Feynman graphs $G$ whose subgraphs $G_{\{1,2,3\}}$ with colors $\{0, 1, 2, 3\}$ and~$G_{\{4,5,6\}}$ with colors $\{0, 4, 5, 6\}$ are both melonic, then it is clear that they are the graphs which maximize the number of faces at f\/ixed number of bubbles.
Gurau's degree formula for~$G_{\{1,2,3\}}$ and~$G_{\{4,5,6\}}$ gives
\begin{gather*}
F_{01}(G) + F_{02}(G) + F_{03}(G) = 2(p(B)-1) b(G) + 3 - \omega(G_{\{1,2,3\}}),\\
F_{04}(G) + F_{05}(G) + F_{06}(G) = 2(p(B)-1) b(G) + 3 - \omega(G_{\{4,5,6\}}),
\end{gather*}
and therefore by summing both equations
\begin{gather*}
\sum_{c=1}^6 F_{0c}(G) - 4(p(B)-1) b(G) = 6 - \omega(G_{\{1,2,3\}}) - \omega(G_{\{4,5,6\}}) \leq 6.
\end{gather*}
We can write the left-hand side in the same form as $\delta_{s_B}(G)$,
\begin{gather*}
\sum_{c=1}^6 F_{0c}(G) - 4(p(B)-1) b(G) = \sum_{c=1}^6 F_{0c}(G) - 5 E(G) + (p(B) + 4) b(G),
\end{gather*}
from which we conclude that the enhancement for any bubble of this type is $s_B = p(B) + 4$. If ${\mathcal G}(B)$ contains graphs for which $\omega(G_{\{1,2,3\}}) = \omega(G_{\{4,5,6\}}) = 0$, then the large~$N$ limit is non-trivial.

The reason we could easily work out the above example is the partition of $B$ into two subgraphs. To formalize this situation, consider a bubble $B$ and a partition of the set of colors
\begin{gather*}
\{1, \dotsc, d\} = \bigcup_{k=1}^L \lambda_k,
\end{gather*}
such that:
\begin{enumerate}\itemsep=0pt
\item \label{item:Cardinality} Each $\lambda_k$, $k=1, \dotsc, L$, has cardinality at least two, $|\lambda_k| \geq 2$,
\item \label{item:Connectivity} The subgraph of $B$ with colors in $\lambda_k$, denoted $B_{\lambda_k}$, is connected.

This assumption can actually be relaxed but the analysis is then more involved~\cite{New1/N}. Notice then that the analysis which follows applies to any bubble. It may however not be conclusive as far as the enhancement is concerned.
\item Assuming that assumption \ref{item:Connectivity} holds, we can further assume that $B_{\lambda_k}$ has an enhance\-ment~$s_{\lambda_k}$. If assumption \ref{item:Connectivity} is relaxed, one can always take $s_{\lambda_k} = |\lambda_k| - 1$ as done originally in~\cite{New1/N} to ensure the existence of the $1/N$-expansion. One could also try and adapt the analysis of~\cite{New1/N} in case $B_{\lambda_k}$ is not connected but each connected piece has a given enhancement.
\end{enumerate}

Let $G\in {\mathcal G}(B)$. We denote $G_{\lambda_k}$ the subgraph of $G$ with all edges of colors $0$ and the edges with colors in $\lambda_k$. It is connected because we have assumed that the restriction of~$B$ to~$\lambda_k$ is. Then the existence of the enhancement $s_{\lambda_k}$ ensures that the number of faces with colors $(0c)$ for $c\in \lambda_k$ can be written like in~\eqref{Power},
\begin{gather*}
\sum_{c\in\lambda_k} F_{0c}(G) = \bigl((|\lambda_k|-1) p(B) - s_{\lambda_k}\bigr) b(G) + \delta_{s_{\lambda_k}}(G_{\lambda_k}),
\end{gather*}
and is bounded, with $d_{k}$ being the maximum of the power of $G_{\lambda_k}$,
\begin{gather*}
\delta_{s_{\lambda_k}}(G_{\lambda_k}) \leq d_k.
\end{gather*}
Summing over the parts of the partition, we get the following counting of the number of faces of~$G$,
\begin{gather*}
\sum_{c=1}^d F_{0c}(G) = \sum_{k=1}^L \sum_{c\in\lambda_k} F_{0c}(G) = \left((d - L) p(B) - \sum_{k=1}^L s_{\lambda_k} \right) b(G) + \sum_{k=1}^L \delta_{s_{\lambda_k}}(G_{\lambda_k}),
\end{gather*}
where we have used $\sum\limits_{k=1}^L |\lambda_k| = d$. We now consider the power of $G$ written for a yet-to-be-determined enhancement $s_B$,
\begin{gather*}
\delta_{s_B}(G) = \sum_{c=1}^d F_{0c}(G) - \bigl((d - 1) p(B) - s_B\bigr) b(G)\\
 \hphantom{\delta_{s_B}(G)}{} = \left( s_B - (L-1)p(B) - \sum_{k=1}^L s_{\lambda_k}\right) b(G) + \sum_{k=1}^L \delta_{s_{\lambda_k}}(G_{\lambda_k}).
\end{gather*}
A $1/N$-expansion exists provided the right-hand side is bounded for any $b(G)$. Since the last sum is bounded, $\sum\limits_{k=1}^L \delta_{s_{\lambda_k}}(G_{\lambda_k}) \leq \sum\limits_{k=1}^L d_{k}$, this imposes $s_B \leq (L-1) p(B) + \sum\limits_{k=1}^L s_{\lambda_k}$. Moreover, a~non-trivial large~$N$ limit exists if and only if~${\mathcal G}(B)$ contains an inf\/inite family of graphs such that $\forall\, k=1, \dotsc, L$, $\delta_{s_{\lambda_k}}(G_{\lambda_k}) = d_k$, and $s_B = (L-1) p(B) + \sum\limits_{k=1}^L s_{\lambda_k}$. This way, we can f\/ind the enhancement of~$B$ from those of its sub-bubbles $B_{\lambda_k}$.

It is still interesting even if the enhancements of the sub-bubbles are not known. Then, taking $s_{\lambda_k} = |\lambda_k| - 1$ ensures the existence of a~$1/N$-expansion, by Gurau's degree theorem, and the bound on the power of the subgraphs is $d_k = |\lambda_k|$. This reduces the new enhancement to $s_B = (L-1) p(B) + d - L$, and the condition $\delta_{s_{\lambda_k}}(G_{\lambda_k}) = d_k$ becomes the vanishing of Gurau's degree for each subgraph, $\omega(G_{\lambda_k}) = 0$.

For $L=1$, one then recovers $s_B = d-1$. For $L\geq 2$, one notices a limitation right away, since it is the same limitation as for $L=1$: the analysis is inconclusive if there are no graphs such that $\omega(G_{\lambda_k}) = 0$. It can nevertheless solve tensor models with non-melonic bubbles, although the large $N$ limit, if non-trivial, is always found to be either similar to the melonic one or planar. This comes from the fact that $\omega(G_{\lambda_k}) = 0$ enforces $G_{\lambda_k}$ to be either melonic if $\lambda_k$ has cardinality greater than 2, or planar if $\lambda_k$ has cardinality 2.

We have thus found the following theorem
\begin{Theorem}
Let $B$ be a bubble, $(\lambda_1, \dotsc, \lambda_L)$ a partition of $\{1, \dotsc, d\}$ such that the assumptions~{\rm \ref{item:Cardinality}} and {\rm \ref{item:Connectivity}} hold. Then the free energy
\begin{gather*}
f_B(t_B) = \ln \int dT d\overline{T}\, \exp \bigl({-}N^{d-1} T\cdot \overline{T} - N^{s_B} t_B P_B(T, \overline{T}) \bigr),
\end{gather*}
with $s_B = d - L + (L-1) p(B)$, admits a $1/N$-expansion. Its large $N$ limit is non-trivial if and only if~${\mathcal G}(B)$ contains an infinite number of graphs satisfying $\omega(G_{\lambda_k}) = 0$ for all $k = 1, \dotsc, L$.

Moreover, if the sub-bubbles $B_{\lambda_k}$ with colors in~$\lambda_k$ are connected and have enhancements $s_{\lambda_k}$, then~$B$ has the enhancement
\begin{gather*}
s_B = (L-1) p(B) + \sum_{k=1}^L s_{\lambda_k}.
\end{gather*}
\end{Theorem}

More details can be found in \cite{New1/N}, in particular the extension to the case where the assumption \ref{item:Connectivity} is dropped. Interestingly, that framework makes it possible to def\/ine tensor models for ``rectangular'' tensors, i.e., where the tensor indices whose positions are in $\lambda_k$ have a range $N_k$, where $N_1, \dotsc, N_L$ can be chosen independently.

We close this section with a simple application of the theorem. At $d=5$, consider the bubble
\begin{gather*}
B = \begin{array}{@{}c@{}} \includegraphics[scale=.6]{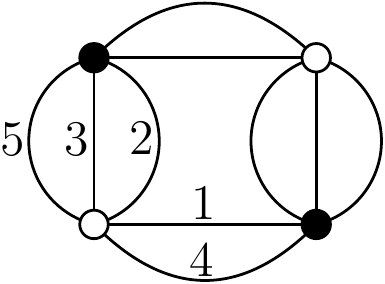} \end{array}
\end{gather*}
With $L=2$, $\lambda_1 = \{1, 2, 3\}$, $\lambda_2 = \{4, 5\}$, one f\/inds the enhancement $s_B = 5>d-1=4$ and the large $N$ limit is non-trivial. Indeed, the sub-bubble with colors $1$, $2$, $3$ is melonic, hence the large~$N$ limit enforces the subgraphs $G_{\{1,2,3\}}$ to be melonic. Moreover, $\omega(G_{\{4,5\}}) = 0$ imposes~$G_{\{4,5\}}$ to be planar. This is in fact automatically the case if $G_{\{1,2,3\}}$ is melonic. Indeed, all graphs contributing to the free energy at large $N$ must have 2-point functions connecting the vertices of the bubble which are connected by the colors~$2$,~$3$ (this is the universality Theorem~\ref{thm:Universality} applied to~$G_{\{1,2,3\}}$). It comes that there are inf\/initely many graphs satisfying $\omega(G_{\lambda_1}) = \omega(G_{\lambda_2}) = 0$, and
\begin{gather*}
\langle P_B(T, \overline{T}) \rangle = N^2 G_2(t_B)^2, \qquad \text{with} \quad G_2(t_B) = \frac1{N^2} \langle T\cdot \overline{T} \rangle.
\end{gather*}
Moreover, from the Schwinger--Dyson equation and the above expectation, one f\/inds that the large $N$ 2-point function satisf\/ies the equation $1-G_2(t_B) - 2t_B G_2(t_B)^2 = 0$, just like the 2-point function of a model with a quartic melonic bubble (see \eqref{MelonicCounting}).

\section{Stuf\/fed Walsh maps} \label{sec:StuffedMaps}

Finding the enhancement of a bubble requires to study the growth of the number of faces of the graphs in ${\mathcal G}(B)$ with the number of bubbles. There are combinatorial objects for which faces are under control: combinatorial maps. Hence, a bijection between ${\mathcal G}(B)$ and maps could be useful. Such a bijection was introduced in \cite{StuffedWalshMaps}.

In the case of quartic bubbles (bubbles with four vertices), the bijection can be seen directly in terms of integrals: one performs a Hubbard--Stratonovich transformation of the quartic tensor integral, which leads to a matrix model with a logarithmic potential \cite{BorelQuartic,BeyondPerturbation, IntermediateT4}. This Hubbard--Stratonovich transformation translates to a bijection between the graphs of the Feynman expansions of both sides of the transformation. Moreover, in the $d=2$ matrix case, the bijection obtained this way is Tutte's bijection between bipartite quadrangulations (which in the dual are maps with vertices of degree four, generated by a~quartic matrix integral) and generic maps (generated by a matrix model with a logarithmic potential). Therefore, the bijection we present can be thought of as an extension of Tutte's bijection to higher dimensions and arbitrary bubbles.

The bijection relies on a repeated use of the following idea. Consider a cyclically ordered list of objects $l = (o_1, \dotsc, o_k)$. It can be represented graphically as an oriented cycle where the objects $o_1, \dotsc, o_k$ are drawn as vertices. Equivalently, one can turn the cycle into a star-shaped map: still representing the objects as vertices, one add an extra vertex $V$ connected to each oject $o_n$ via an edge $e_n$. The cyclic order between the objects thus translates into a cyclic order of the edges, $(e_1, \dotsc, e_k)$, around $V$. Using the counter-clockwise orientation of the plane, the transformation looks like
\begin{gather} \label{CycleToMap}
\begin{array}{@{}c@{}} \includegraphics[scale=.7]{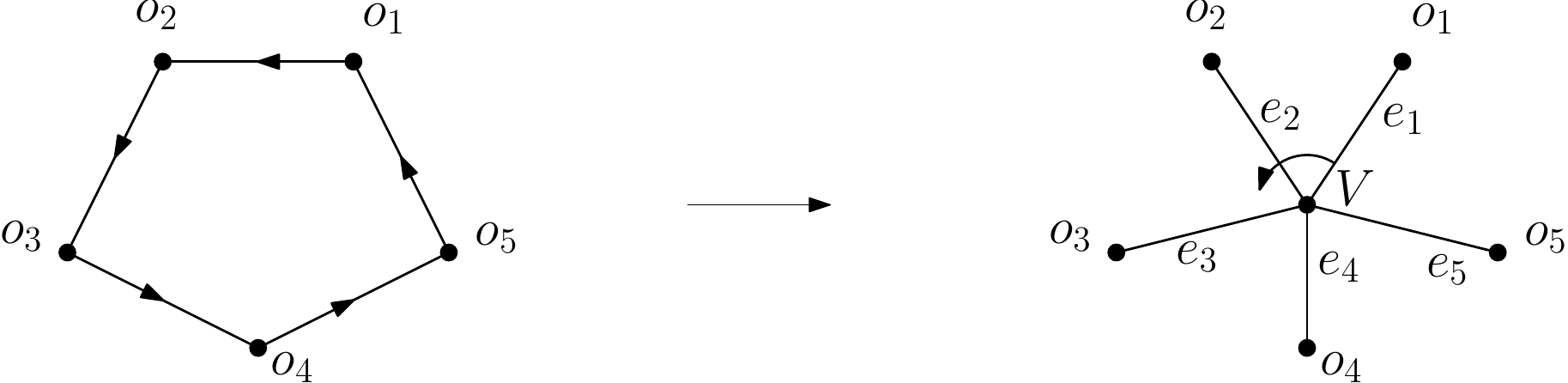} \end{array}
\end{gather}
Crucially, the edge from $o_n$ to $o_{n+1}$ becomes a piece of a face (called a broken face), i.e., a path which goes from $o_n$ to $o_{n+1}$ by following the corner between $e_n$ and $e_{n+1}$ at $V$ counter-clockwise.

\subsection{Representing a bubble as a map}

One starts with a {\it pairing} $\pi$ on $B$, i.e., a partition of its vertices into pairs of black and white vertices. We will map~$B$ to a map $M(B, \pi)$ with blue vertices representing the pairs of~$B$. First, we erase the edges which connect the two vertices of a pair (obviously, one can f\/ind back the missing edges since they are those whose colors are missing at each vertex) and orient the remaining edges of $B$ from their white to black vertices. We can then merge the black and white vertices of each pair into a blue vertex,
\begin{gather*}
\begin{array}{@{}c@{}} \includegraphics[scale=.7]{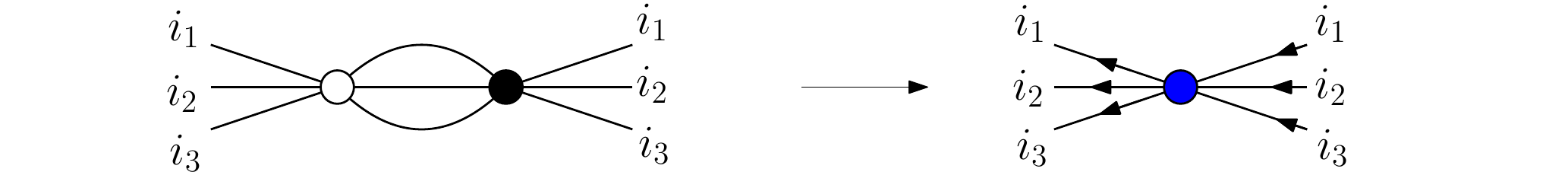} \end{array}
\end{gather*}
and the property ``being incident to a black (white) vertex'' is replaced with ``being ingoing (outgoing) on a blue vertex''. We obtain a graph denoted~$B_\pi$. Thus, if a blue vertex has an ingoing edge of color $j$, it also has an outgoing edge of the same color. Moreover, it cannot have more than two edges of the same color. It other words, the restriction of~$B_\pi$ to a f\/ixed color is a disjoint union of oriented cycles. For instance,
\begin{gather} \label{ExampleBubbleBijection}
B = \begin{array}{@{}c@{}} \includegraphics[scale=.5]{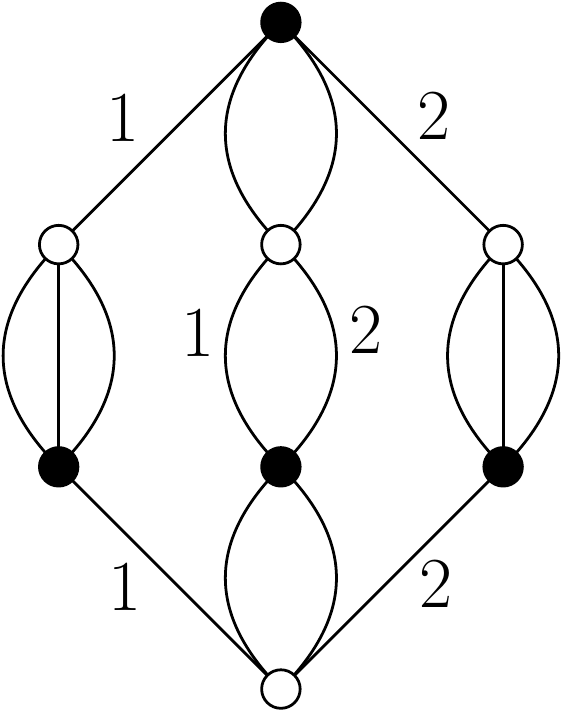} \end{array} \overset{\pi}{\to} \begin{array}{@{}c@{}} \includegraphics[scale=.5]{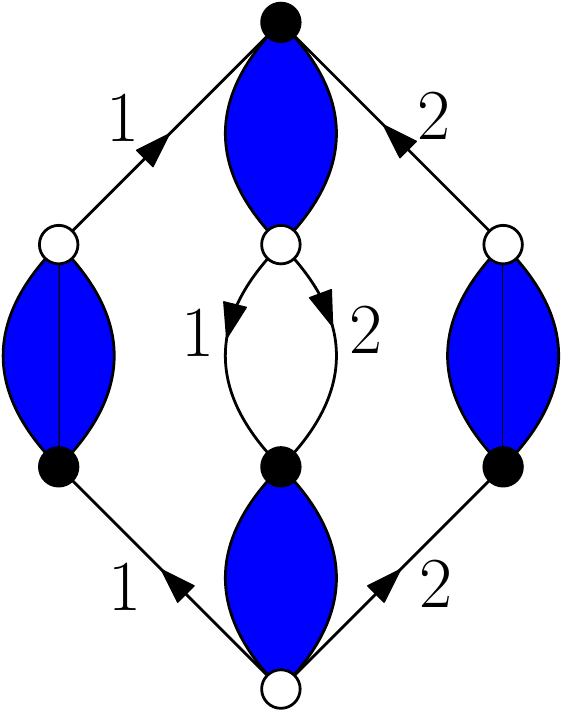} \end{array} \to B_\pi = \begin{array}{@{}c@{}} \includegraphics[scale=.5]{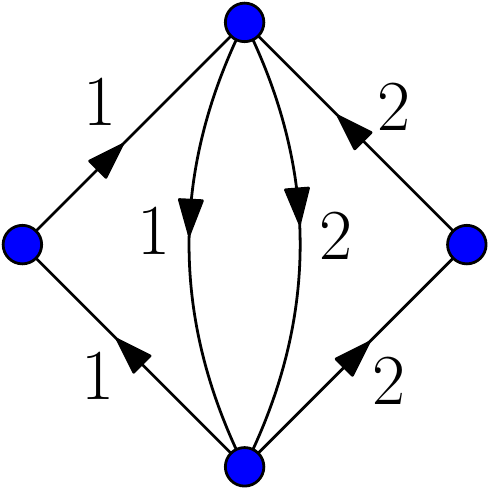} \end{array}
\end{gather}

Therefore, in this section, the word \emph{cycle} means equivalently a simple closed path of edges of a f\/ixed color in $B_\pi$ or a cyclic list which alternates pairs of vertices of $B$ (represented by the blue vertices of $B_\pi$) and edges of a f\/ixed color (and visiting them at most once).

Let $l^{(j)} = (\rho_1, \dotsc, \rho_{k_j})$ be such a cycle of color $j$ which goes along the (cyclically ordered) blue vertices $\rho_1, \dotsc, \rho_{k_j}$. We apply the transformation~\eqref{CycleToMap} to~$l^{(j)}$: we add a box-vertex~$V_{l^{(j)}}$ and connect it with edges of color~$j$ to the vertices $\rho_1, \dotsc, \rho_{k_j}$. This is done so that the (cyclic) counter-clockwise order around~$V_{l^{(j)}}$ is the same as the (cyclic) order of the blue vertices around~$l^{(j)}$. This transformation is performed on all cycles of all colors in~$B_\pi$. Continuing the example of~\eqref{ExampleBubbleBijection},
\begin{gather} \label{ExampleBubbleBijectionContinued}
B_\pi = \begin{array}{@{}c@{}} \includegraphics[scale=.5]{ExampleBubbleBijection3} \end{array} \to M(B, \pi) = \begin{array}{@{}c@{}} \includegraphics[scale=.5]{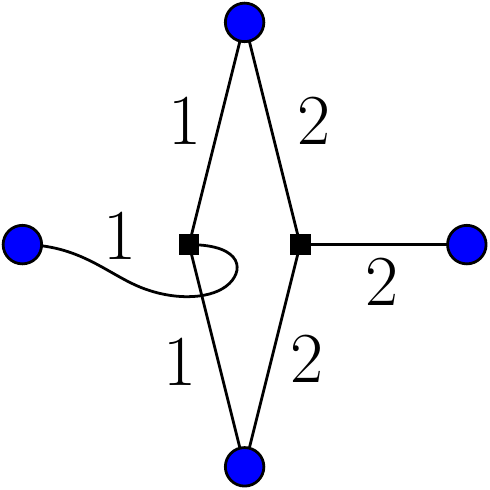} \end{array}.
\end{gather}
The order between dif\/ferent colors incident on a blue vertex is irrelevant and can be chosen arbitrarily.

Clearly, an edge of color $j\in \{1, \dotsc, d\}$ is either not represented in $M(B, \pi)$ if it connects two vertices of a pair, or it is represented in $B_\pi$ by an oriented edge between two blue vertices. In the latter case, it is then represented in $M(B, \pi)$ as a counter-clockwise corner at a box-vertex between two edges of color~$j$,
\begin{gather*}
\begin{array}{@{}c@{}} \includegraphics[scale=.7]{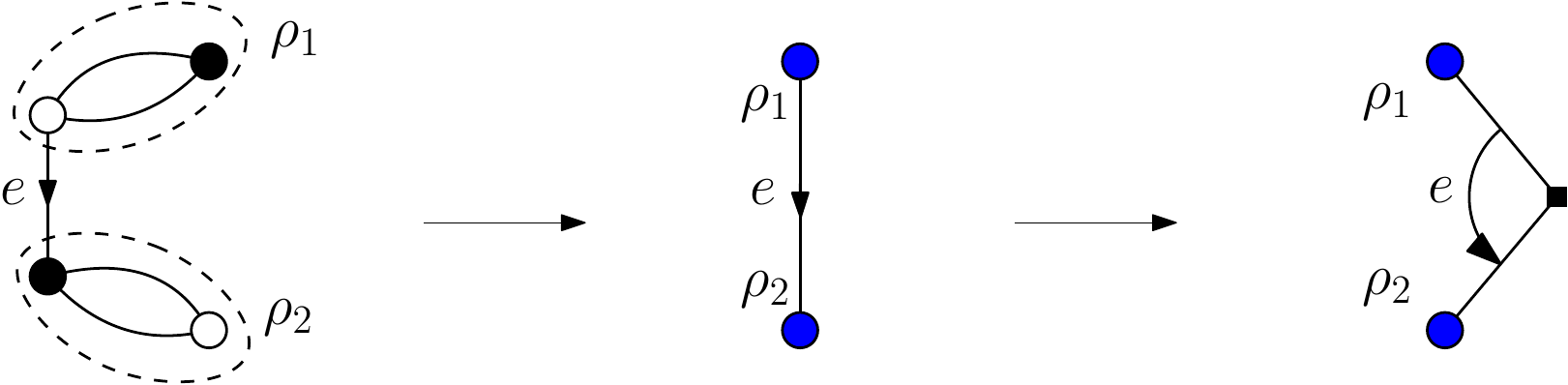} \end{array}
\end{gather*}

\subsection{The universal part of the bijection}

We now need to connect the maps $M(B, \pi)$ together. In the original graph $G\in {\mathcal G}(B)$, bubbles are connected by edges of color $0$. The pairing $\pi$ on $B$ induces a pairing $\pi_G$ on all the graphs $G\in {\mathcal G}(B)$ by choosing $\pi$ on all the copies of $B$ contained in $G$.

The pairing $\pi_G$ can be thought of as a permutation: given a labeling of the vertices, it maps each white vertex to a black vertex. Similarly, the edges of color~$0$ can be encoded through a~permuta\-tion~$\tau_0$: using this same labeling, $\tau_0$ maps a black vertex to a white vertex if there is an edge of color~$0$ between them.

We now consider the cycles of $\pi_G \circ \tau_0$. Graphically, start from a black vertex and follow the incident edge of color~$0$ to a white vertex (this is $\tau_0$). This white vertex is part of a pair given by~$\pi$, so jump to the black vertex of that pair, and repeat. The cycles obtained that way contain all edges of color~$0$ and all pairs.

In this section, the word \emph{cycle} thus means equivalently a cycle of a permutation or a cyclic list of objects in a colored graph which alternates edges of color 0 and pairs of vertices from~$\pi$ (and visiting them at most once).

Denote $c = (\rho_1, \dotsc, \rho_k)$ a cycle of $\pi_G \circ \tau_0$ which encounters the pairs $\rho_1, \dotsc, \rho_k$. If the vertices of~$\rho_j$ are connected by the colors of a subset ${\mathcal I}_j \subset \{1, \dotsc, d\}$, denote $\hat{{\mathcal I}_j} = \{1, \dotsc, d\}\setminus {\mathcal I}_j$ its complement. These are the colors of the edges that, when being followed from a vertex of~$\rho_j$ lead to other pairs in dif\/ferent cycles of $\pi_G \circ \tau_0$.

We represent each cycle $c = (\rho_1, \dotsc, \rho_k)$ by a black vertex $v_c$ of degree $k$. A pair $\rho_j$ is represented in the map $M(B, \pi)$ by a blue vertex. We therefore connect $v_c$ to the blue vertices representing $\rho_1, \dotsc, \rho_k$. An edge connecting $v_c$ to $\rho_j$ is decorated with the color set $\hat{{\mathcal I}}_j$. Moreover, one has to record the order in which the pairs are encountered along $c$. This gives rise to a cyclic order of the edges incident to $v_c$. This is summarized graphically as follows
\begin{gather*}
\begin{array}{@{}c@{}}
\includegraphics[scale=.6]{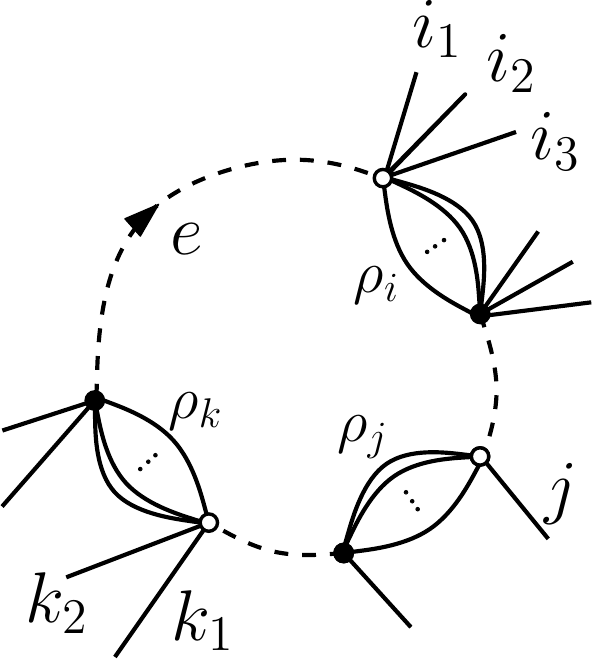}
\end{array}
\qquad \to \qquad
\begin{array}{@{}c@{}}
\includegraphics[scale=.7]{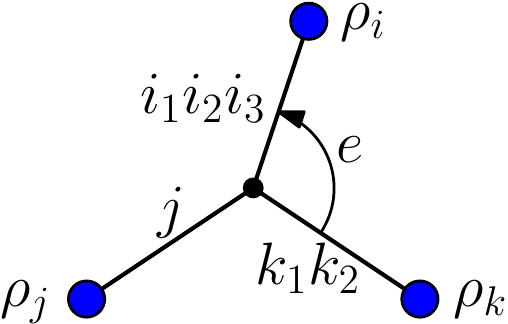}
\end{array}
\end{gather*}
Notice that the edges of color $0$ are mapped to corners. It is done in such a way that following an edge of color $0$ from its black vertex to its white vertex corresponds to the counter-clockwise orientation of the corner. Therefore, an oriented path which alternates edges of color~$0$ (from black to white vertices) and edges of another f\/ixed color, say $j\in\{1, \dotsc, d\}$, along the cycle becomes a path which follows corners counter-clockwise until an edge carrying the color $j$ is met. It is a closed path (i.e., a face of colors $(0j)$ of $G$) if and only if the color $j$ is not incident to the vertex. In other words, the parts of the faces of colors~$(0j)$ which go through such a cycle become the corners around the vertex between edges which carry the color $j$.

Notice that this part is {\it universal}: it does not depend on $B$. In principle, all subsets of $\{1, \dotsc, d\}$ are allowed on the edges connecting the blue and black vertices. The restriction to a~particular bubble only puts restrictions on the subsets of colors which can ef\/fectively appear.

\subsection{Bijection with stuf\/fed Walsh maps}

The set of maps obtained by gluing copies of $M(B, \pi)$ via black vertices is denoted ${\mathcal W}(B,\pi)$. The maps $W\in {\mathcal W}(B, \pi)$ are called stuf\/fed Walsh maps, because if one replaces each copy of~$M(B, \pi)$ contained in $W$ with a white vertex, one would have a bipartite map where the white vertices can be thought of as hyper-edges (this is Walsh's representation of hypermaps~\cite{Walsh}). In fact, if~$B$ is a cycle (closed path alternating edges of colors 1 and 2), then the bijection precisely leads to bipartite maps which represent hypermaps~\cite{StuffedWalshMaps}. However a white vertex is generically not a faithful representation of $M(B,\pi)$, so a map $W$ can be thought as obtained from stuf\/f\/ing a~bipartite map with copies of $M(B, \pi)$.

Due to the stuf\/f\/ing, a map $W\in {\mathcal W}(B, \pi)$ has three types of vertices: blue ones represent pairs, box-vertices incident to edges of color $j$ represent cycles of pairs and edges of color $j$, and black vertices represent cycles of pairs and edges of color~$0$.

We introduce $W^{(c)}$ the submap of $W$ with all the edges whose color sets contain $c\in\{1, \dotsc, d\}$. It is typically disconnected and isolated vertices are taken into account.

\begin{Theorem}
For any choice of pairing $\pi$ there is a bijection between ${\mathcal G}(B)$ and ${\mathcal W}(B, \pi)$ which maps copies of $B$ to copies of $M(B, \pi)$, pairs of vertices to blue vertices, edges of color $j\in\{1, \dotsc, d\}$ to corners around box-vertices between edges of color $j$, edges of color $0$ to corners around black vertices while the faces of colors~$(0c)$ are mapped to the faces of $W^{(c)}$.
\end{Theorem}

The only bit we have not explained yet is the fact that the faces of colors $(0c)$ become the faces of the submap $W^{(c)}$. All the construction pointed that way. Indeed, an edge of color~$c$ from its white vertex to its black vertex is mapped to a counter-clockwise corner at a box-vertex. Moreover, an edge of color $0$ from its black vertex to its white vertex is mapped to a counter-clockwise corner at a black vertex of $W$ between two edges containing the color $c$. Following edges of colors~$c$ and~$0$ therefore amounts to following corners counter-clockwise between edges which contain the color $j$, as shown below
\begin{gather*}
\begin{array}{@{}c@{}} \includegraphics[scale=.5]{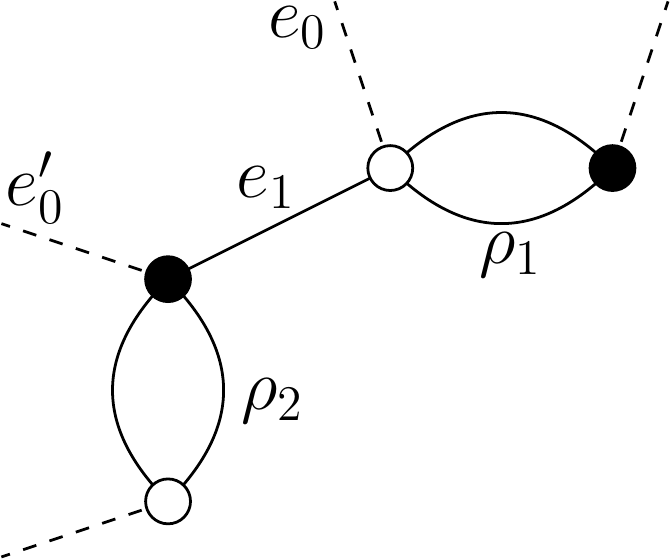} \end{array} \qquad \to \qquad \begin{array}{@{}c@{}} \includegraphics[scale=.5]{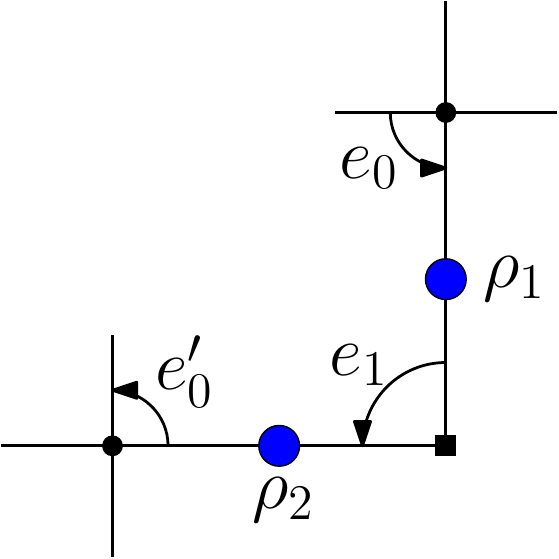} \end{array}
\end{gather*}
Notice that no order is specif\/ied at the blue vertices because for a f\/ixed color, a blue vertex is always bivalent (and the order between distinct colors is irrelevant).

\subsection{Projected maps and trees}

{\sloppy Stuf\/fed Walsh maps remain complicated objects to study due to the structure of the map~$M(B, \pi)$ which represent the bubble $B$. As such, the maps~$M(B, \pi)$ encodes all the richness of the possible bubbles. Yet, for any $B$, there are some tree-like maps for which we can easily f\/ind the number of faces.

}

To see that, we introduce the notion of projected maps. For a stuf\/fed Walsh map $W\in{\mathcal W}(B, \pi)$, its {\it projected map} $PW$ is def\/ined by representing all submaps $M(B, \pi)$ as vertices, say white vertices, while preserving the rest of $W$. This requires to f\/ix an ordering of the edges around these white vertices. As a consequence, $PW$ loses the structure of the bubble $B$ and becomes an edge-colored hypermap.

\begin{Proposition}
The number of faces of a map $W$ whose projected map $PW$ is a tree is
\begin{gather} \label{FacesProjectedTrees}
F(W) = (F(B^\pi) - d)V_\circ(W) + d,
\end{gather}
where $B^\pi$ is the map which consists in a single $M(B, \pi)$ whose blue vertices are each connected to a univalent black vertex, and $V_\circ(W)$ is the number of copies of $M(B, \pi)$ in $W$.
\end{Proposition}

\begin{figure}[t]
\centering\includegraphics[scale=.4]{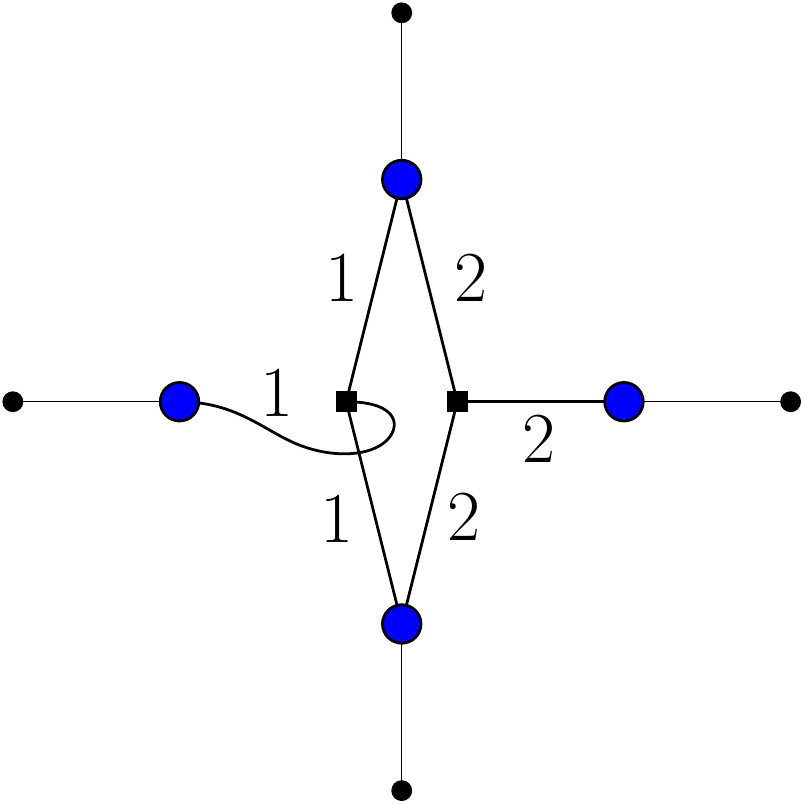} \qquad\qquad \includegraphics[scale=.4]{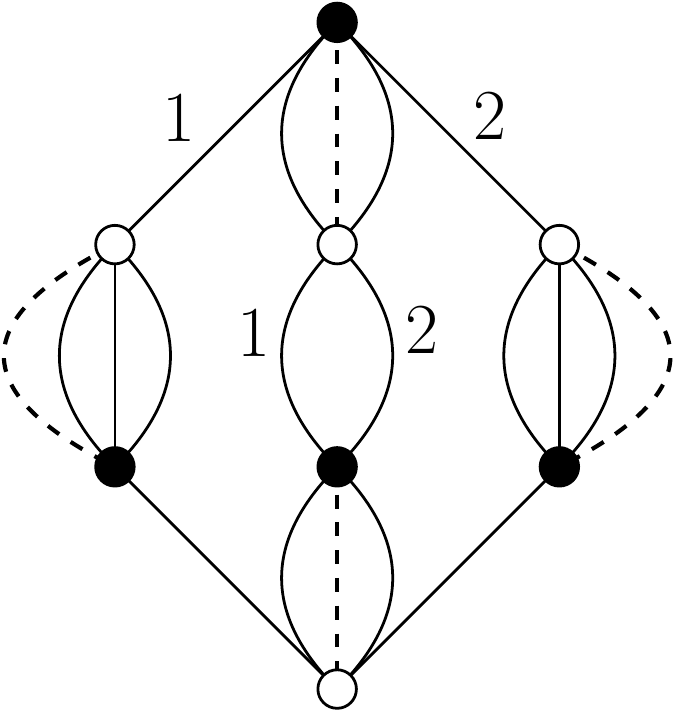}
\caption{We use the bubble and pairing of~\eqref{ExampleBubbleBijection}. On the left is the map $B^\pi$ obtained by connecting the blue vertices of $M(B, \pi)$ (taken from \eqref{ExampleBubbleBijectionContinued}) to univalent black vertices. On the right is the corresponding colored graph, with dashed edges for the color 0. It simply consists in adding the color 0 between the vertices of each pair of~$\pi$. Indeed, blue vertices represent pairs of vertices of $B$, and connecting them to univalent black vertices correspond to closing cycles made of a pair and an edge of color~0.}\label{fig:ExampleGaussianContraction}
\end{figure}

An example of the map $B^\pi$ is provided in Fig.~\ref{fig:ExampleGaussianContraction} in the case the bubble $B$ and the pairing are those of equation~\eqref{ExampleBubbleBijection}.

In other words, the number of faces of such maps grows linearly with the number of sub\-maps~$M(B, \pi)$, with a known coef\/f\/icient. This coef\/f\/icient features $B^\pi$ which is easily interpreted in terms of colored graphs. Indeed, it is the graph obtained by representing the pairing $\pi$ with edges of color $0$ connecting the two vertices of each pair in $B$. This is a contribution to the expectation of the polynomial $P_B(T, \overline{T})$ in the Gaussian distribution.

Let us \emph{assume} that $B$ and $\pi$ are such that the number of faces of any map $F(W)$ is bounded by the number of faces of those whose projected maps are trees, at f\/ixed $V_\circ$. We can then f\/ind the enhancement $s_B$,
\begin{gather} \label{EnhancementProjectedTree}
s_B = d + (d-1)p(B) - F(B^\pi).
\end{gather}
Indeed, since the bijection preserves the number of faces, $F(W) = F(G)$ and since $V_\circ(W) = b(G)$, our assumption means that there exists a family of graphs $G\in {\mathcal G}(B)$ which maximizes the number of faces at f\/ixed number of bubbles. Hence $F(G) \leq (F(B^\pi) - d) b(G) + d$. Then the power $\delta(G)$ of a graph $G$ is bounded like
\begin{gather*}
\delta_{s_B}(G) \leq d - \bigl[(d-1) p(B) - (F(B^\pi) - d) - s_B\bigr] b(G).
\end{gather*}
Choosing $s_B$ as in \eqref{EnhancementProjectedTree} is then the only choice which provides a bound independent of the number of bubbles and at the same time which can be saturated by an inf\/inite family of graphs: those in bijection with maps whose projected maps are trees.

The dif\/f\/icult point is to prove the assumption that the number of faces is bounded by \eqref{FacesProjectedTrees} for some choice of $\pi$. It has been proved in some cases \cite{MelonoPlanar, StuffedWalshMaps}, and no counter-examples have been found yet. Notice that in most cases where it was proved, maps whose projected maps are trees are not the only maps which maximize the number of faces. Often, the set of such maps include a subset of planar maps.

It is important to keep in mind that the graphs represented by trees in \eqref{FacesProjectedTrees} depend on the choice of $\pi$. Any choice of $\pi$ is a valid one. However, depending on what one plans to use the bijection for, some choices can be more convenient and useful. In our case, we are interested in f\/inding the maps which maximize the number of faces. Therefore, we would like $\pi$ to help us characterize the maps which maximize the number of faces and make them as simple as possible, like trees for instance (or rather maps whose projected maps are trees).

Dif\/ferent pairings $\pi$ on a f\/ixed bubble $B$ can lead to projected maps which are the same trees, but with a dif\/ferent counting of faces (due to dif\/ferent values of $F(B^\pi)$). For instance, if $B$ is melonic, there is a canonical pairing $\pi$ (the one which maximizes $F(B^\pi)$). If chosen, this pairing is such that the number of faces is indeed bounded by the number of faces of maps which project on trees: maps which maximize the number of faces are those whose projected maps are trees. In fact, they are precisely the melonic graphs built from~$B$. However, for a~dif\/ferent choice of pairing, our bijection does not map melonic graphs to trees. Instead, (maps whose projected maps are) trees then correspond to an entirely dif\/ferent family of graphs which certainly do not maximize the number of faces.

From this example, it appears that a convenient choice of $\pi$ is one such that $F(B^\pi)$ is maximal, say~$\pi_*$. Only then one might hope to prove that (maps whose projected maps are) trees are dominant. In fact, it is proved in~\cite{StuffedWalshMaps} that maps whose projected maps have a single cycle (in the sense of a simple closed path of edges in the projected map) always have fewer faces than trees, provided one chooses $\pi_*$ as pairing for the bijection. For other choices of $\pi$, there is no conjecture at all on what the maps which maximize the number of faces look like.

\section{The quartic case} \label{sec:Quartic}

\subsection{Description of the dominant maps} \label{sec:QuarticDominant}

The bijection we have presented simplif\/ies quite a bit in the quartic case, i.e., the case of bubbles with four vertices (as the bijection is then generated by a Hubbard--Stratonovich transformation). Indeed, those bubbles have two pairs of vertices which are mapped in $M(B, \pi)$ to two blue vertices. They are connected by the colors in ${\mathcal I} = \{i_1, \dotsc, i_k\}$. The box-vertices for each of those colors are then bivalent, meaning that they can be transformed to just edges between the two blue vertices. Since one gets such an edge for all colors which connect the two pairs, the map $M(B, \pi)$ can simply be turned to an edge between two blue vertices, decorated with the colors which connect the two pairs,
\begin{gather*}
\begin{array}{@{}c@{}} \includegraphics[scale=.65]{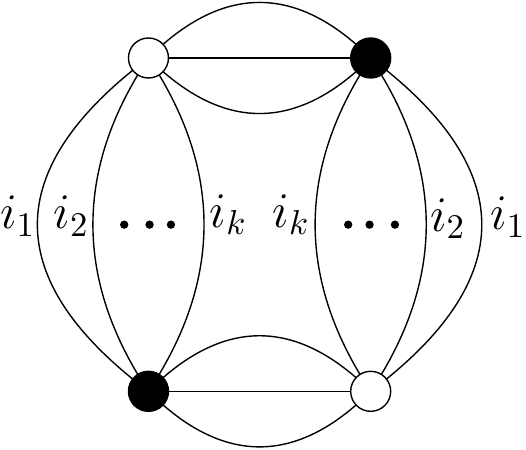} \end{array} \qquad \to \qquad \begin{array}{@{}c@{}} \includegraphics[scale=.65]{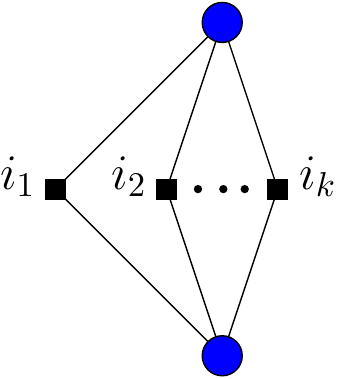} \end{array} \qquad \to \qquad \begin{array}{@{}c@{}} \includegraphics[scale=.65]{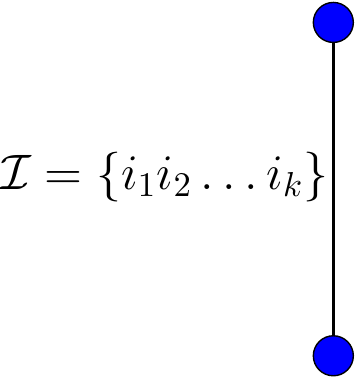} \end{array}
\end{gather*}
Each blue vertex therefore connects to another blue vertex and a black vertex, with edges carrying the same color set ${\mathcal I}$. They are thus bivalent and can be erased. One ends up with maps with only black vertices of arbitrary degrees, and edges colored by ${\mathcal I}$.

When several types of quartic bubbles are allowed, the generalization is obvious: each quartic bubble becomes an edge of a map with a color set. We will consider quartic melonic and necklace bubbles with the following pairings
\begin{gather*}
\begin{array}{@{}c@{}} \includegraphics[scale=.5]{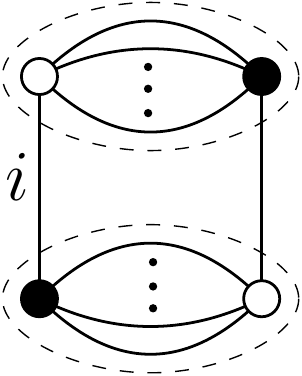} \end{array} \qquad \to \qquad \begin{array}{@{}c@{}} \includegraphics[scale=.5]{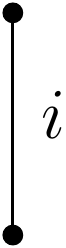} \end{array},\hspace{2cm}
\begin{array}{@{}c@{}} \includegraphics[scale=.5]{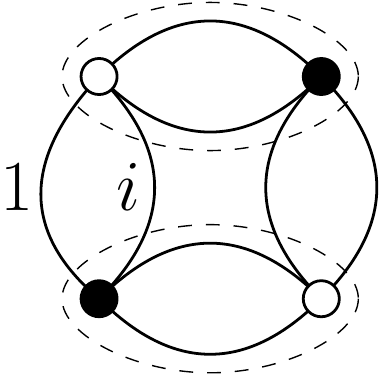} \end{array} \qquad \to \qquad \begin{array}{@{}c@{}} \includegraphics[scale=.5]{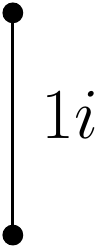} \end{array}
\end{gather*}

We already know the enhancements for all those bubbles, $s=d-1$ for the melonic ones and $s=4$ for the quartic necklaces at $d=4$. It remains to f\/ind the maps which maximize the number of faces when the bubbles are all used together. The results have appeared in \cite{MelonoPlanar}. The monocolored edges have to be bridges (or cut-edges, i.e., their removals disconnect the map). We can therefore temporarily assume that $M$ is a map without monocolored edge. It therefore only has bicolored edges with colors $(1i)$. To f\/ind the maps which maximize the number of faces, we use a result of \cite{StuffedWalshMaps} which gives the dif\/ference between the number of faces of a stuf\/fed Walsh map and the number of faces of another one whose projected map is a tree with the same number of copies of $M(B,\pi)$. In our context, maps and projected maps are the same, so we f\/ind that the dif\/ference between the number of faces of $M$ and that of a tree $T$ with the same number of edges is
\begin{gather} \label{FacesQuarticModel}
F(M) - F(T) = -4 l(M) + 2 \sum_{i=1}^4 l(M^{(i)}) - 2 \sum_{i=1}^4 g\big(M^{(i)}\big).
\end{gather}
Here $l(M) = E(M) - V(M) + 1$ is the cyclomatic number of $M$, $M^{(i)}$ is the (typically non-connected) submap obtained by keeping only the edges whose color set contains the color~$i$, and~$l(M^{(i)})$,~$g(M^{(i)})$ are its cyclomatic number and genus. Notice that the color~$1$ labels all edges, therefore $l(M) = l(M^{(1)})$. Since $M^{(2)}$, $M^{(3)}$, $M^{(4)}$ are edge-disjoint submaps, it comes
\begin{gather} \label{CyclesQuarticModel}
l\big(M^{(2)}\big) + l\big(M^{(3)}\big) + l\big(M^{(4)}\big) \leq l(M).
\end{gather}
The right-hand side of \eqref{FacesQuarticModel} is thus nonpositive and therefore $F(M) \leq F(T)$.

To get the equality $F(M) = F(T)$, $i)$ the map $M$ should have all its colored submaps planar\footnote{In general, $M^{(c)}$ is not connected. Its genus is the sum of the genera of its connected components.}, $g(M^{(c)}) = 0$ for $c=1, 2, 3, 4$, and~$ii)$ the equality has to hold in~\eqref{CyclesQuarticModel}. The latter condition means that the edges of every cycle (i.e., simple closed path of edges) have the same color type, either~$(12)$ or~$(13)$ or~$(14)$.

\begin{figure}[t]
\centering\includegraphics[scale=.4]{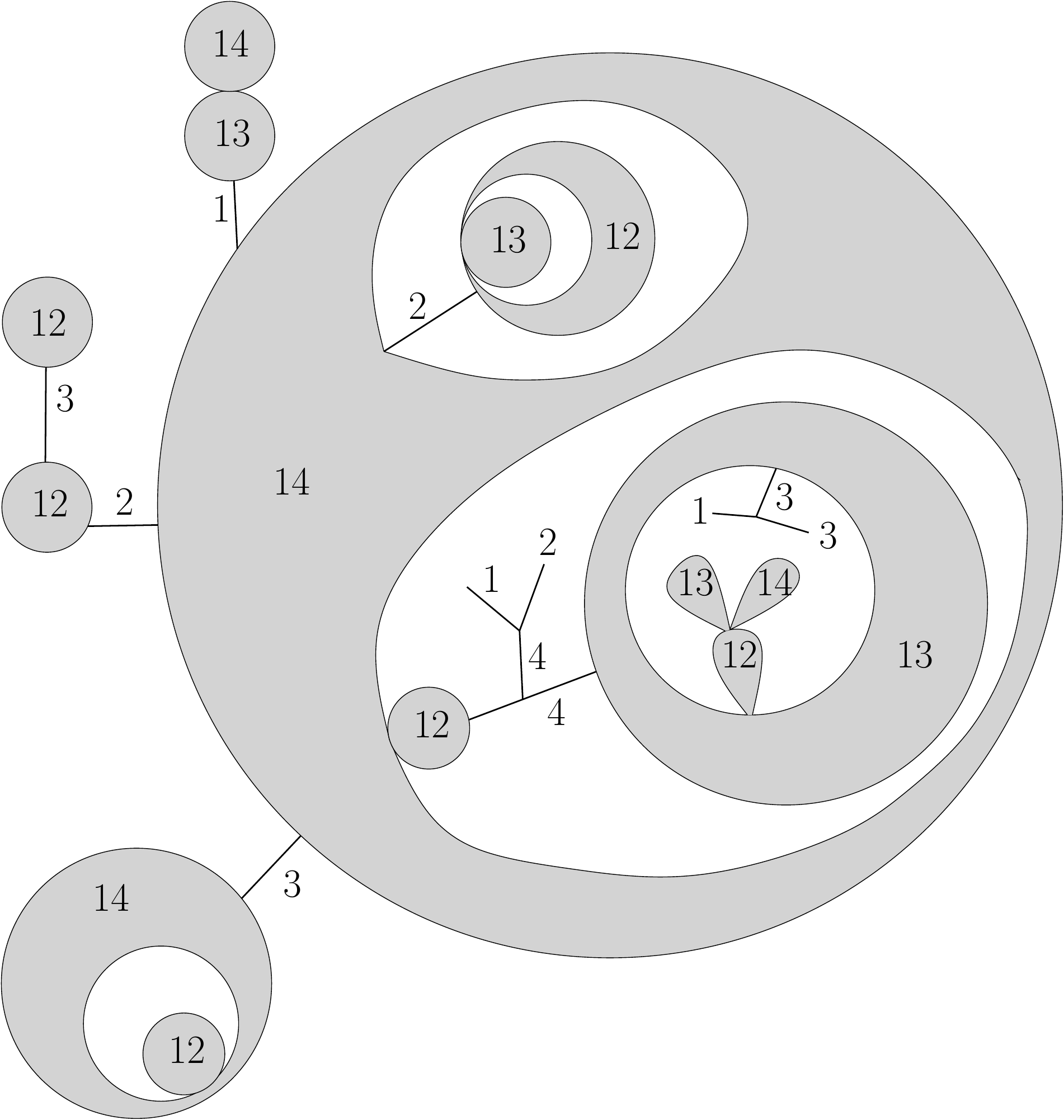}
\caption{This is a schematic example of a graph in ${\mathcal G}^{(3, 4)}_{\max}$. The greyed regions with labels~$(1c)$ are arbitrary planar connected components with edges of type $(1c)$. Every two such regions can only meet on cut-vertices, leading to a (nested) cactus-like structure.}\label{fig:Cactus}
\end{figure}

Let ${\mathcal G}_{\max}^{(k, q)}$ be the set of colored graphs made of $q$ types of quartic melonic bubbles and $k$ types of quartic necklace bubbles which maximize the number of faces at f\/ixed number of bubbles. It is mapped to the set ${\mathcal M}_{\max}^{(k,q)}$ of maps which satisfy the following conditions,
\begin{itemize}\itemsep=0pt
\item vertices have unbounded degrees,
\item monocolored edges can have $q$ possible colors (as we allow $q\leq d$ types of quartic melonic bubbles),
\item monocolored edges are bridges,
\item bicolored edges carry the color type $(1c)$ where $c$ can take $k$ values (and $k\leq 3$ since there are at most the types $(12)$, $(13)$ and $(14)$),
\item the submaps $M^{(c)}$ made of all edges which carry the color $c$ are planar,
\item every cycle has a f\/ixed color type $(1c)$.
\end{itemize}
An schematic example is given in Fig.~\ref{fig:Cactus}. The constraint that every cycle has a f\/ixed color type is equivalent to the following: Every two connected submaps of dif\/ferent color types can only meet at a cut-vertex (recall that a cut-vertex is a vertex whose removal increases the number of connected components of the map). Indeed, assume that there is a cycle made of edges of several color types. We can choose as submaps the chains of edges of the same color type and maximal length along the cycle and those submaps clearly meet on vertices which are not cut-vertices. The other way around, assume that there exist two connected submaps $M'$, $M''$ of dif\/ferent color types which meet on a vertex $v$ which is not a cut-vertex. For all vertices $u\in M'$, $w\in M''$ there is a path from~$u$ to~$w$ going through~$v$. Since~$v$ is not a cut-vertex, the map remains connected after its removal, and therefore there exists a path between $u$ and $w$ which do not go through~$v$. These two paths can be concatenated to form a cycle (removing the vertices which are visited more than once) which has at least two distinct color types.

\subsection{Enumeration of the dominant maps} \label{sec:Enumeration}

We denote $\bar{{\mathcal M}}_{\max}^{(k,q)}$ the set of rooted maps (i.e., with an marked oriented edge) satisfying all those conditions, and its generating function
\begin{gather*}
f_{(k,q)}(t,\lambda) = \sum_{M\in\bar{{\mathcal M}}_{\max}^{(k,q)}} t^{E(M)} \lambda^{E_m(M)},
\end{gather*}
where $E(M)$ is the total number of edges and $E_m(M)$ the number of monocolored edges of $M$. This is actually the 2-point function of the quartic tensor model. Indeed, the latter is a sum over colored graphs with a marked edge of color~$0$. A marked edge of color~$0$ is incident to the black vertex of a bubble. In the map point of view, this bubble is an edge~$e$ and the edge of color~$0$ is a corner incident to~$e$. We can then orient $e$ outward, and the construction can be reversed, i.e., an oriented edge of a map is equivalent to a marked edge of color~$0$. Moreover, since quartic bubbles have two indistinguishable white (or black) vertices, there are two equivalent ways to glue a bubble to a graph. This leads to $C(G) = 2^{b(G)} = 2^{E(M)}$ which is being re-absorbed in~$t$.

Notice that the only constraint on monocolored edges is that every one of them is a bridge, so their colors are irrelevant. It means that one can set a unique color to monocolored edges and simply rescale $\lambda$ by $q$, $f_{(k,q)}(t, \lambda) = f_{(k,1)}(t, q\lambda)$. We thus forget about~$q$ and study $f_k(t,\lambda) \equiv f_{(k,q=1)}(t,\lambda)$.

We consider the map consisting of a single vertex as a rooted map so that $f_k(t,\lambda)$ starts with~$1$. All other terms have at least one edge. Assume that the root edge $e$ is monocolored, then as it is a bridge, it separates two connected components which can be canonically rooted on the f\/irst edge after $e$ by going counter-clockwise around each vertex of $e$. Those two connected components are thus in $\bar{{\mathcal M}}_{\max}^{(k,1)}$. We thus have
\begin{gather} \label{TreeContributions}
f_k(t, \lambda) = 1 + t \lambda f_k(t, \lambda)^2 + \text{maps rooted on a bicolored edge}.
\end{gather}
The only dif\/f\/icult part of the enumeration is to take into account the constraint that every cycle has a single color type. To do so, we notice that connected components consisting of edges of color type $(1c)$ and $(1c')$ for $c\neq c'$ can only touch at a cut-vertex which separate them (i.e., removing the vertex separates the two connected components).

Let $M\in \bar{{\mathcal M}}_{\max}^{(k,1)}$ rooted on a bicolored edge $e$ and let $M'$ be the maximal non-separable submap containing $e$. As explained in \cite{GouldenJacksonBook}, $M$ can be recovered from $M'$ by inserting rooted maps $M_\alpha \in \bar{{\mathcal M}}_{\max}^{(k,1)}$ on each corner of $M'$ (if $M_\alpha$ is inserted in the counter-clockwise corner between $e_1$ and $e_2$, one can root $M_\alpha$ on the f\/irst edge met after $e_1$). This decomposition holds for generic planar maps \cite{GouldenJacksonBook}. In our case, the key point is that $M'$ has a single color type (the same as its root edge $e$). However, the corner insertions $M_\alpha$ can be arbitrary and in particular rooted on an edge of arbitrary color type.

Let $P(t)$ be the generating function of non-separable rooted planar maps counted with respect to the number of edges. Since the number of corners is twice the number of edges, one can consider that there are two maps $M_\alpha, M_\beta$ inserted on $M'$ for each edge. The contribution of the maps rooted on a bicolored edge is thus $P(t f_{k}(t,\lambda)^2)-1$ for a single color type ($P$ contains the single vertex map, which we have already taken into account in~\eqref{TreeContributions}, hence the $-1$). Adding the $k$ types of bicolored root edge to~\eqref{TreeContributions} therefore gives
\begin{gather} \label{Decomposition}
f_k(t, \lambda) = 1 - k + t \lambda f_k(t, \lambda)^2 + k P\bigl(t f_{k}(t,\lambda)^2\bigr).
\end{gather}
It remains to describe the generating function $P(x)$. It satisf\/ies the following algebraic sys\-tem~\cite{GouldenJacksonBook} (more about where it comes from below)
\begin{gather} \label{ParametricNonSeparable}
 x = u (1-u)^2,\qquad P = (1-u)(1+3u).
\end{gather}
The latter can be turned into an algebraic system describing $f_k(t, \lambda)$. Indeed, by setting $x = tf^2$, one directly gets a parametrization of $P(t f^2)$. In particular, $t f^2 = u (1-u)^2$. Moreover from~\eqref{Decomposition} we also know that $f = 1 - k + \lambda x + k P(x)$. By plugging $x = u (1-u)^2$ and $P(x) = (1-u)(1+3u)$, the following system is obtained
\begin{gather} \label{AlgebraicSystem}
t f^2 = u (1-u)^2, \qquad f = k (1-u)(1+3u) - k + 1 + \lambda u(1-u)^2,
\end{gather}
and $f_k(t, \lambda)$ is the solution of this system for $f$ after elimination of $u$. One then gets the polynomial equation
\begin{gather}
t f_k(t, \lambda)^2 \Bigl(2 k^3 + k^2 (\lambda - 18) + 3\lambda - 4\lambda k + \bigl(18 k^2 - 6\lambda + 4 \lambda k\bigr) f_k(t, \lambda)\nonumber\\
\qquad{} + \bigl(3\lambda (1+t\lambda) - 27 k^3 t - 18 k^2 t \lambda - 2 k t \lambda^2\bigr) f_k(t, \lambda)^2
 - 3 t \lambda^2 f_k(t, \lambda)^3 + t^2 \lambda^3 f_k(t, \lambda)^4 \Bigr)\nonumber\\
\qquad{} - \bigl(f_k(t, \lambda) - 1\bigr) \bigl(f_k(t, \lambda) + k - 1\bigr)^2 = 0.\label{PolynomialEq}
\end{gather}
Notice that when allowing for a single type of bicolored edges, i.e., $k=1$, and no monocolored edges, i.e., $\lambda=0$, one should be enumerating generic planar maps. The polynomial equa\-tion~\eqref{PolynomialEq} indeed reduces to the well-known quadratic equation on the generating function of rooted planar maps,
\begin{gather} \label{MapsGF}
27 t^2 A(t)^2 + (1 - 18t) A(t) + 16t - 1 = 0,
\end{gather}
with $A(t) = f_{k=1}(t, \lambda = 0)$. In fact,~\cite{GouldenJacksonBook} proceeds precisely the other way around to get the system~\eqref{ParametricNonSeparable}. Indeed, as is standard, one can f\/ind~\eqref{MapsGF} independently, using Tutte's equation for instance. It can thus be shown that $A(t)$ satisf\/ies the system $t = \alpha (1-3\alpha)$, $A = (1-4\alpha)(1-3\alpha)^{-2}$. This system combined with the decomposition~\eqref{Decomposition} in the case $k=1$, $\lambda = 0$, i.e., $A(t) = P(t A(t)^2)$, leads to~\eqref{ParametricNonSeparable} after some change of variables~\cite{GouldenJacksonBook}. What we are doing here is thus extend $A(t) = P(t A(t)^2)$ to~\eqref{Decomposition} and use~\eqref{ParametricNonSeparable} to get to~\eqref{AlgebraicSystem}.

We are going to discuss the singularities of $f_k(t, \lambda)$ with respect to $t$. Since $t$ counts the total number of edges, those singularities describe the asymptotic behavior of the number of maps in~$\bar{{\mathcal M}}^{(k,1)}_{\max}$ with respect to the number of edges. We f\/ind it more convenient to work with the algebraic system~\eqref{AlgebraicSystem} rather than the polynomial equation~\eqref{PolynomialEq}.

The system \eqref{AlgebraicSystem} describes solutions for $f$ and $u$ as functions of $t$, with parameters~$k$,~$\lambda$. The system becomes singular with respect to $t$ when the Jacobian vanishes. The dif\/ferential system reads
\begin{gather*}
\begin{pmatrix} 2t f & -(1-u)(1-3u)\\ 1 & -2k(1-3u) - \lambda (1-u)(1-3u) \end{pmatrix} \begin{pmatrix} f'\\ u'\end{pmatrix} = \begin{pmatrix} -f^2 \\ 0\end{pmatrix},
\end{gather*}
where the prime indicates the derivative with respect to $t$. The singularities are thus determined by adding the vanishing of the determinant of the above matrix to the system \eqref{AlgebraicSystem},
\begin{gather}
t f^2 = u (1-u)^2,\nonumber \\
 f = k (1-u)(1+3u) - k + 1 + \lambda u(1-u)^2,\nonumber \\
 (1 - 3u) \bigl(1 - u - (2k + \lambda (1-u))2 t f\bigr) = 0.\label{CriticalSystem}
\end{gather}
Solutions to this system can be given using radicals but the explicit expressions are lengthy and not particularly suitable for detailed analysis. Instead we solve explicitly the case where $k=1$, i.e., a single type of bicolored edges allowed, and the case $\lambda = 0$, i.e., no monocolored edges (which can only be bridges). Those two situations will both give the same phase diagram, which we conjecture extends to the case of generic $(k, \lambda)$:
\begin{itemize}\itemsep=0pt
\item For $k$ and $\lambda$ small enough, a large typical map consists of several planar components each of a f\/ixed bicolored type, connected by a f\/inite number of monocolored edges (with weight $\lambda$) and/or cut-vertices which can connect two components of dif\/ferent color types. Therefore the criticality is expected to be that of planar maps, with a singularity $(t_c - t)^{3/2}$.

\item For $k$ or $\lambda$ large enough, a large typical map has mostly monocolored edges (which form trees) and/or numerous cut-vertices separating planar components of dif\/ferent color types (when $k$ gets larger, the probability of adding a component of a dif\/ferent color type increases). The planar components of f\/ixed color type remain non-critical. The maps are thus dominated by branching processes, so a singularity $(t_c - t)^{1/2}$ is expected, i.e., the universality class of trees.
\item Between those two phases, there should be a regime where bicolored planar components are in inf\/inite number (connected by inf\/initely many monocolored edges and/or cut-vertices) and each of them becomes inf\/inite too. This phase thus has a proliferation of baby universes and the expected singularity is $(t_c - t)^{2/3}$.
\end{itemize}

The proliferation of baby universes is a known phenomenon, which has been encountered in the context of multi-trace matrix models in \cite{AlvarezBarbon, BarbonDemeterfi,Das, KlebanovHashimoto, Korchemsky}. A multi-trace matrix model has as interaction products of matrix traces, like $\operatorname{Tr} M^{k_1} \dotsm \operatorname{Tr} M^{k_q}$ for the matrix~$M$. An interaction of this type can be thought of as the superposition of $q$ vertices of degrees $k_1, \dotsc, k_q$. For maps which maximize the number of faces, the planar components which contain each of those~$q$ vertices can only touch at the point where they are superimposed. This leads to planar components connected in a cactus way. This phenomenon is analogous to the fact that planar components of dif\/ferent color types can only touch at vertices which separate them in our model. This is therefore a branching process also similar to connecting planar components via bridges. This analogy explains the presence of the proliferation of baby universes in tensor models.

We can further f\/ind an equation on the radius of convergence of $f_k(t, \lambda)$ in terms of $k$, $\lambda$ in the regime where it has a square-root singularity. Let us look at solutions of \eqref{PolynomialEq} with a square-root singularity, $f_k(t, \lambda) = a + b \sqrt{\rho - t} + o(\sqrt{\rho - t})$. At orders 0 and 1 in $\sqrt{\rho - t}$, equation \eqref{PolynomialEq} gives algebraic relations between $a, \rho, k, \lambda$. Eliminating $a$, the resultant of those two equations then reduces to
\begin{gather*}
0 = -\lambda\rho + \big(4 k^3+40 \lambda k^2+12 \lambda ^2-8 \lambda ^2 k\big) \rho \nonumber\\
\hphantom{0 =}{}+ \big({-}128 k^5-192 \lambda k^4-96 \lambda ^2 k^3-320 \lambda k^3-48 \lambda ^3-16 \lambda ^3 k^2-32 \lambda ^2 k^2+64 \lambda ^3 k\big)\rho^2 \nonumber\\
\hphantom{0 =}{}+ \big(1024 k^6+2048 \lambda k^5+1536 \lambda ^2 k^4+64 \lambda ^4+512 \lambda ^3 k^3-512 \lambda ^2 k^3+64 \lambda ^4 k^2\\
\hphantom{0 =}{}-512 \lambda ^3 k^2-128 \lambda ^4 k\big)\rho^3.
\end{gather*}
Anticipating the cases studied below, notice that it has the solution $\rho = \frac{\lambda}{4(1+\lambda)^2}$ for $k=1$ and $\rho = \frac{k + \sqrt{k(k-1)}}{16 k^2}$ for $\lambda = 0$.

By further developing some ansatz of the form $f_k(t, \lambda) = a_0 + a_1 (\rho - t)^\alpha + a_2 (\rho - t) + \cdots$ for some well-chosen exponents~$\alpha$, and plugging them into~\eqref{PolynomialEq}, one might try and match the coef\/f\/icients to f\/ind algebraic relations between them. Instead of pursuing that approach, we will describe in details two cases where the values of the exponent $\alpha$ can be found directly.

\subsubsection{A single type of necklace bubble}

Allowing for a single type of necklace bubbles corresponds to allowing for a single type of bicolored edges, i.e., $k=1$. This case has in fact been solved in \cite{MelonoPlanar} with the so-called ``trees of necklaces'' bubbles which generalize the present case $k=1$ beyond quartic bubbles. It was solved in \cite{MelonoPlanar} using the Schwinger--Dyson equations of tensor models which were shown to reduce to the loop equations of multi-trace matrix models, already studied in \cite{AlvarezBarbon, BarbonDemeterfi,Das, KlebanovHashimoto, Korchemsky}. Here we present another approach, based on the system \eqref{AlgebraicSystem} instead, which to our knowledge has not appeared in the matrix model literature.

Notice that the polynomial equation \eqref{PolynomialEq} reduces to a quartic equation,
\begin{gather*}
1 - 16 t - \bigl(1 + 2 t(\lambda - 9)\bigr) f_{k=1}(t, \lambda) + \bigl(3\lambda t + t^2 (\lambda^2 - 18 \lambda - 27)\bigr) f_{k=1}(t, \lambda)^2\\
\qquad{} - 3 t^2 \lambda^2 f_{k=1}(t, \lambda)^3 + t^3 \lambda^3 f_{k=1}(t, \lambda)^4 = 0,
\end{gather*}
which itself reduces to the quadratic equation \eqref{MapsGF} for the generating function of planar maps $A(t)$ by setting $\lambda = 0$.

\begin{figure}[t]
\centering\includegraphics[scale=.9]{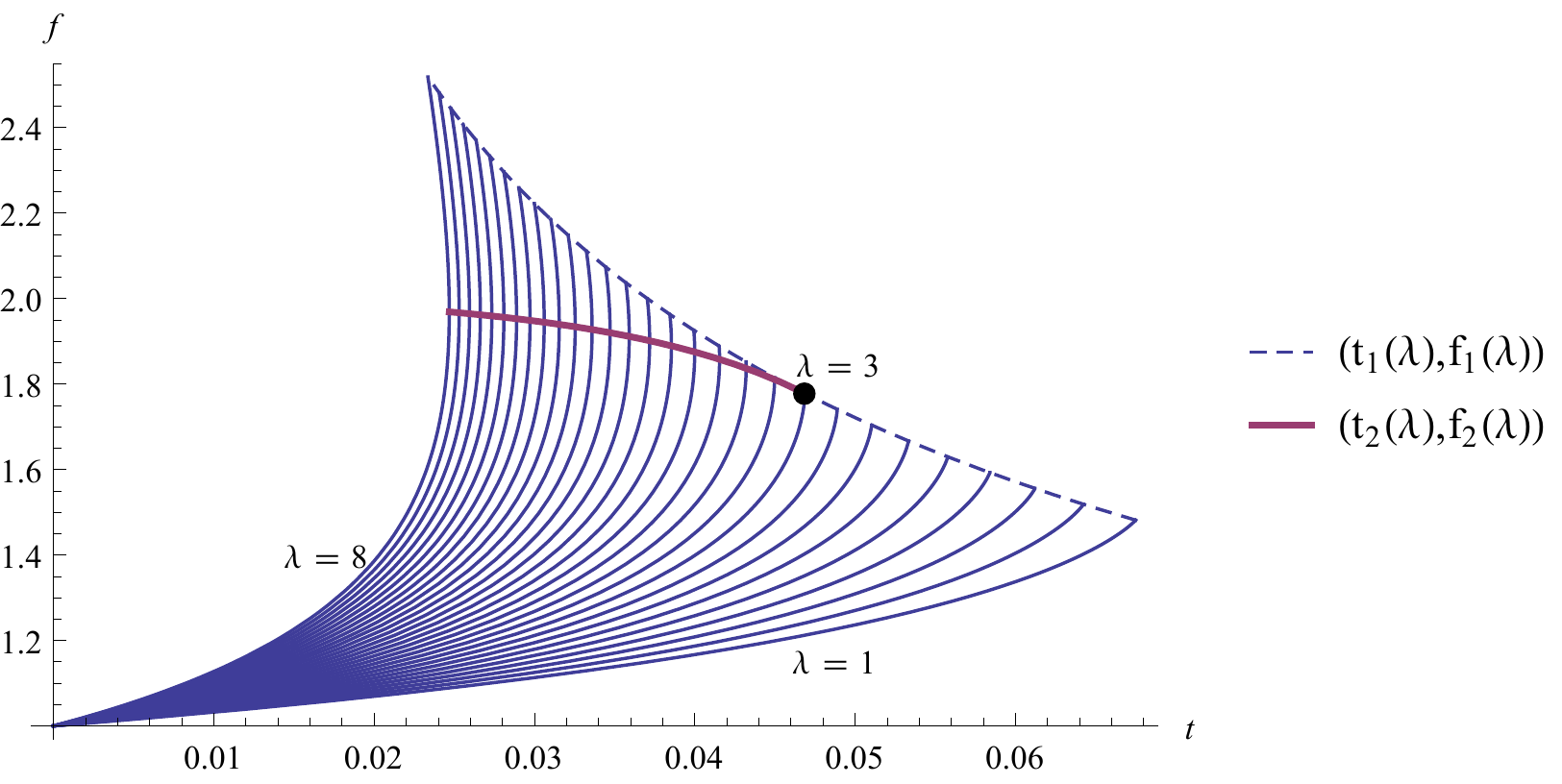}
\caption{The curves represent $f$ as a function of $t$ for $\lambda = 1$ up to $\lambda = 8$ with a step of~$.25$. The dashed line is the singular locus for $\lambda \leq 3$, and the thick line for $\lambda \geq 3$ where a $f(t)$ develops a vertical tangent.}\label{fig:SingleNecklacePlot}
\end{figure}

The system \eqref{CriticalSystem} then admits the following solutions (depending on $\lambda$)
\begin{gather*}
u_1(\lambda) = \frac{1}{3},\qquad f_1(\lambda) = \frac{4}{27}(\lambda + 9), \qquad t_1(\lambda) = \frac{27}{4 (\lambda + 9)^2}
\end{gather*}
which is the singular locus for $\lambda \leq 3$, and
\begin{gather*}
u_2(\lambda) = \frac{1}{\lambda},\qquad f_2(\lambda) = 2 \frac{\lambda^2 - 1}{\lambda^2}, \qquad t_2(\lambda) = \frac{\lambda}{4 (1 + \lambda)^2}
\end{gather*}
for $\lambda \geq 3$. The function $f_{k=1}(t, \lambda)$ is represented in Fig.~\ref{fig:SingleNecklacePlot} as a function of~$t$ for various values of~$\lambda$. For $\lambda \leq 3$, $f(t)$ hits a singularity at~$t_1(\lambda)$. For $\lambda \geq 3$, the f\/irst singularity encountered while coming from $(t=0, f=1)$ becomes the one at $(t_2(\lambda), f_2(\lambda))$.

Performing some Taylor expansion of the system \eqref{AlgebraicSystem} around the above critical points, one f\/inds the singular behaviors. For $\lambda<3$, one f\/inds
\begin{gather*}
f_{k=1}(t, \lambda) = f_1(\lambda) + \frac{16 (\lambda + 3)(\lambda + 9)^3}{729 (\lambda - 3)}(t_1(\lambda) - t)
 + \frac{64 (\lambda + 9)^{11/2}}{6561 (3-\lambda)^{5/2}} (t_1(\lambda) - t)^{3/2} \\
\hphantom{f_{k=1}(t, \lambda) =}{} + o\bigl( (t_1(\lambda) - t)^{3/2} \bigr).
\end{gather*}
It thus extends the known singularity of the generating function of planar maps,
\begin{gather} \label{MapsSingularity}
A(t) = \frac{4}{3} - 16 \left(\frac{1}{12} - t\right) + 64 \sqrt{3} \left(\frac{1}{12} - t\right)^{3/2} + o\left(\left(\frac{1}{12} - t\right)^{3/2} \right),
\end{gather}
with $t_1(\lambda = 0) = 1/12, f_1(\lambda = 0) = 4/3$, to the case where planar maps can be connected by bridges with weight $\lambda$.

For $\lambda >3$, one gets
\begin{gather*}
f_{k=1}(t, \lambda) = f_2(\lambda) - \frac{4(1 + \lambda)^2}{\lambda^{5/2}} \sqrt{\lambda^2 - 2\lambda - 3} (t_2(\lambda) - t)^{1/2} + o\bigl( (t_2(\lambda) - t)^{1/2} \bigr),
\end{gather*}
which is the behavior expected for trees. Finally, at $\lambda = 3$, one gets $t = 3/64 - 81/512 (1/3 - u)^3 + o((1/3 - u)^3)$ and $f = 16/9 - 6 (1/3 - u)^2 + o((1/3 - u)^2)$, and therefore
\begin{gather*}
f_{k=1}(t, \lambda = 3) = \frac{16}{9} - \frac{128}{3^{5/3}}\left(\frac{3}{64} - t\right)^{2/3} + o\left( \left(\frac{3}{64} - t\right)^{2/3} \right),
\end{gather*}
which corresponds to the phase where baby universes (each planar submap of f\/ixed color type becoming critical) proliferate because there is an inf\/inity of them connected by bridges in a tree-like fashion. This phase was originally described in the context of multi-trace matrix models~\cite{AlvarezBarbon, BarbonDemeterfi,Das, KlebanovHashimoto, Korchemsky}.

\subsubsection{No monocolored edges}

The maps with no monocolored edges, i.e., $\lambda = 0$, corresponds to the case where the quartic melonic bubbles are not allowed whereas $k$ types of necklace bubbles are. {\it A priori}, $k$ is an integer at most $3$. Notice however that the algebraic system~\eqref{AlgebraicSystem} makes sense for $k$ a positive real number.

Setting $\lambda = 0$, the polynomial equation \eqref{PolynomialEq} becomes of order 4,
\begin{gather*}
\begin{split}
& (k - 1)^2 - (k-1)(k-3) f_k(t) + \big(3 - 2k - 18 k^2 t + 2k^3 t\big) f_k(t)^2 \\
& \qquad{}+ \big(18 k^2 t - 1\big) f_k(t)^3 - 27 k^3 t^2 f_k(t)^4 = 0,
\end{split}
\end{gather*}
where we have set $f_k(t) = f_k(t, \lambda = 0)$. The linear and constant terms vanish at $k=1$, thus reproducing~\eqref{MapsGF}.

\begin{figure}[t]
\centering\includegraphics[scale=.9]{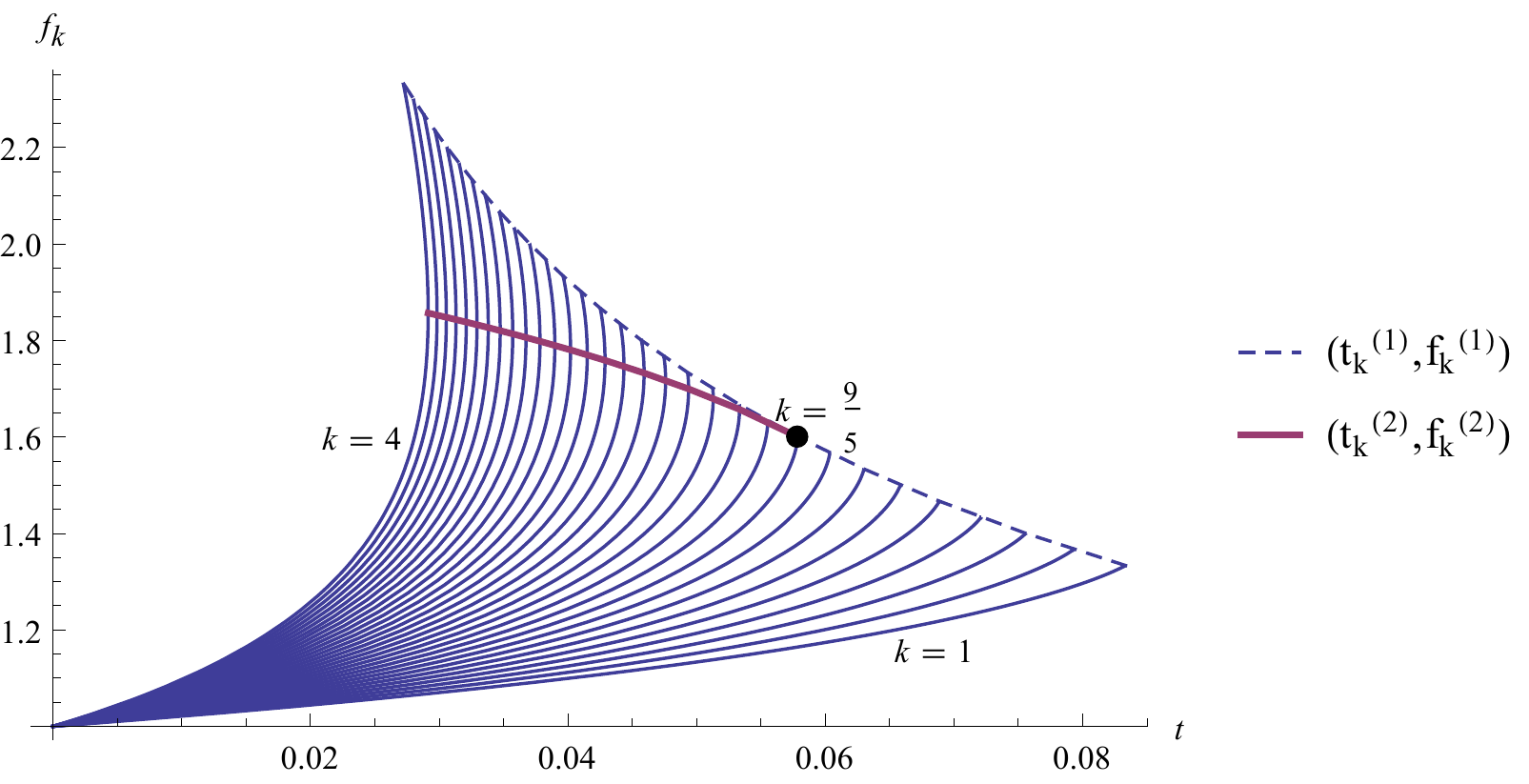}
\caption{The curves represent $f_k$ as a function of~$t$ for $k = 1$ up to $k = 4$ with a step of~$.1$. The dashed line is the singular locus for $k \leq 9/5$, and the thick line for $k \geq 9/5$ where a~$f_k(t)$ develops a~vertical tangent.}\label{fig:NoBridgesPlot}
\end{figure}

The system \eqref{CriticalSystem} which determines the singularities has four solutions at $\lambda=0$. Two of them give negative values to $f_k$ for all $k\geq 1$. We dismiss them and the ``physical'' solutions are the two others. For $k\leq 9/5$ the singularities are on the following line parametrized by $k$
\begin{gather*}
u_k^{(1)} = \frac{1}{3},\qquad f_k^{(1)} = 1 + \frac{k}{3},\qquad t_k^{(1)} = \frac{4}{3 (k+3)^2}
\end{gather*}
while for $k\geq 9/5$, they are on
\begin{gather*}
u_k^{(2)} = 1 - \sqrt{\frac{k-1}{k}},\qquad f_k^{(2)} = 4\bigl(1 - k + \sqrt{k(k-1)}\bigl),\qquad t_k^{(2)} = \frac{k+\sqrt{k(k-1)}}{16\,k^2},
\end{gather*}
and those two solutions coincide at $k=9/5$, as shown on the plot of Fig.~\ref{fig:NoBridgesPlot}.

The behavior of $f_k(t)$ close to the singular points can be obtained by performing Taylor expansions of \eqref{AlgebraicSystem}. For $k<9/5$, that gives
\begin{gather*}
f_k(t) = f_k^{(1)} - \frac{k(k+3)^3}{9 - 5k} \big(t_k^{(1)} - t\big) + \sqrt{3} k \frac{(k + 3)^{11/2}}{(9 - 5k)^{5/2}}\big(t_k^{(1)} - t\big)^{3/2} + o\bigl(\big(t_k^{(1)} - t\big)^{3/2}\bigr),
\end{gather*}
which indeed reproduces \eqref{MapsSingularity} when $k\to 1$. For $k>9/5$, one f\/inds
\begin{gather*}
f_k(t) = f_k^{(2)} + 16 \sqrt{k \bigl(20 k^3 - 31 k^2 + 11 k - \sqrt{k (k-1)} (20 k^2 - 21 k + 3)\bigr)} \big(t_k^{(2)} - t\big)^{1/2} \\
\hphantom{f_k(t) =}{}
+ o\bigl(\big(t_k^{(2)} - t\big)^{1/2}\bigr).
\end{gather*}
meaning that $f_k(t) - f_k^{(2)} \sim \sqrt{t_k^{(2)} - t}$ for some non-zero, explicit constant. As in the case $k=1$ and $\lambda>3$, the universality class is that of ordinary trees. This is interpreted as the fact that when~$k$ is large enough, it becomes very likely that a typical vertex separates planar components of dif\/ferent color types. This produces cacti-like structures similar to trees.

Eventually, for $k=9/5$, the critical values are $f_k^{(1)} = f_k^{(2)} = 8/5$ and $t_k^{(1)} = t_k^{(2)} = 25/432$. Moreover, the singularity takes the form
\begin{gather*}
f_{k=9/5}(t) = \frac{8}{5} - \frac{432}{25 \times 5^{1/3}} \left(\frac{25}{432} - t\right)^{2/3} + o\left(\left(\frac{25}{432} - t\right)^{2/3}\right).
\end{gather*}

\section{Conclusion}

We started the present article by explaining how random tensor models generate discrete, higher-dimensional (pseudo-)manifolds, the same way random matrix models generate combinatorial maps. We insisted on generalizing the notion of $2p$-angulations (or more generally maps with restrictions on the allowed face degrees), by using colored triangulations with prescribed bubbles. A~bubble is a generalization of a $2p$-angle, obtained by gluing colored simplices until the boun\-da\-ry is formed by $(d-1)$-simplices of a f\/ixed color, much like a~$2p$-angle is a gluing of~$2p$ colored triangles with $2p$ edges of color $0$ on its boundary. We then went on studying the set of colored triangulations built by gluing copies of an arbitrarily chosen bubble~$B$, and denoted~${\mathcal G}(B)$. Importantly, those bubbles and colored triangulations can be represented as regular, edge-colored graphs with respectively~$d$ and $d+1$ colors.

The program of random tensors can then be put in purely combinatorial terms. It consists in classifying the graphs of ${\mathcal G}(B)$ with respect to the number of bubbles and the number of faces, where a face is a cycle alternating the colors $0$ and $c$, for $c\in\{1, \dotsc, d\}$. The classif\/ication by Gurau and Schaef\/fer \cite{GurauSchaeffer} does precisely that for the set of all colored graphs, using Gurau's degree $\omega(G) \geq 0$. However, it is believed that for a typical bubble~$B$, the degree $\omega(G)$ for $G\in {\mathcal G}(B)$ scales with the number of bubbles in~$G$. In other words, the generating function at f\/ixed degree is polynomial and no interesting continuum limit can be achieved.

We have called an enhancement of $B$ the maximal value of $s_B$ such that
\begin{gather*}
\delta_{s_B}(G) = F(G) - \bigl[(d-1) p(B) - s_B\bigr] b(G)
\end{gather*}
is bounded for all $G\in{\mathcal G}(B)$. The strategy we used to f\/ind enhancements was to try and f\/ind the graphs which maximize the number of faces at f\/ixed number of bubbles forming the set~${\mathcal G}_{\max}(B)$. We have also shown that f\/inding~$s_B$ is crucial to def\/ining tensor models with non-trivial large~$N$ limits.

Sections \ref{sec:BubbleGluing}, \ref{sec:Slices} and \ref{sec:StuffedMaps} were then devoted to three techniques developed in the literature to f\/ind enhancements, using combinatorial tools. We have moreover taken the opportunity of reviewing those techniques to formalize and actually extend two of them borrowed from \cite{New1/N, MelonoPlanar, DoubleScaling} in Sections~\ref{sec:BubbleGluing} and~\ref{sec:Slices}. Section~\ref{sec:StuffedMaps} then reviews the bijection proposed recently in~\cite{StuffedWalshMaps}.

This bijection is f\/inally used in Section~\ref{sec:Quartic} to perform the enumeration of the graphs in ${\mathcal G}_{\max}(B)$ in the quartic cases (with four-vertex bubbles). The relevant graphs were already described in~\cite{MelonoPlanar} but we have $i)$ revisited their derivation using the new bijection, $ii)$ perform the enumeration beyond ~\cite{MelonoPlanar}.

The set ${\mathcal G}_{\max}(B)$ was described for other bubbles in~\cite{StuffedWalshMaps}, but the set of universality classes that colored triangulations can reach is far from being understood. We know that it includes all universality classes of combinatorial maps, since the necklace bubbles are matrix models in disguise (as shown in Section \ref{sec:NonGaussianExample}). The three techniques presented in this article are fairly recent and have been applied to a handful of bubbles only. It is possible that those techniques, as well as combinations of them, can lead to the enumeration of ${\mathcal G}_{\max}(B)$ for bubbles $B$ which have not been investigated yet. For instance, one might use the bijection to f\/ind new enhancements, then the techniques of Sections \ref{sec:BubbleGluing} and \ref{sec:Slices} to derive other enhancements from them.

While the phase diagram resulting from the enumeration in Section \ref{sec:Quartic} is not really surprising, we think the method is worth discussing. We indeed used a decomposition of the relevant maps onto non-separable planar maps whose generating function is known \cite{GouldenJacksonBook}. In particular, we did not use directly the Schwinger--Dyson/loop/Tutte equations for quartic models (however, we use the system \eqref{ParametricNonSeparable} for non-separable planar maps which is derived in \cite{GouldenJacksonBook} from Tutte's equation for planar maps). It would be interesting to write (using the tensor integrals, or matrix integrals as discussed below) and solve the Schwinger--Dyson equations for quartic models directly, since this is one of the more traditional methods in matrix models and also the root of the topological recursion.

In fact, the maps counted in Section \ref{sec:Quartic} can be generated by a matrix model \cite{MelonoPlanar}. More generally, stuf\/fed Walsh maps based on any bubble, as described in Section \ref{sec:StuffedMaps}, can be generated using random matrix integrals as shown in \cite{StuffedWalshMaps}. This means that there are Schwinger--Dyson equations which characterize the relevant generating functions. We have not however found any way to solve those equations yet. The dif\/f\/iculty lies in the fact that the corresponding matrix models are multi-matrix, multi-trace models, so that the matrices typically do not commute. The set of relevant observables is thus more dif\/f\/icult to analyze and so is the set of Schwinger--Dyson equations.

In the quartic case with only necklace bubbles, the matrix model is as follows. Let $V = \mathbb{R}^N$ be the def\/ining representation space of $N\times N$ matrices. We introduce $N^2\times N^2$ Hermitian matrices $\psi_{12}$, $\psi_{13}$, $\psi_{14}$ and denote $\Psi_{1c}$ the matrix which acts on $V\otimes V\otimes V\otimes V$ as $\psi_{1c}$ on the f\/irst and $c$-th copies of $V$ and as the identity on the two other factors. Then
\begin{gather} \label{NecklaceMatrixModel}
\int d\psi_{12} d\psi_{13} d\psi_{14}\, \exp \left\{{-}N^2 \sum_{c=2}^4 \operatorname{Tr}_{V^{\otimes 2}} \psi_{1c}^2 + \operatorname{Tr}_{V^{\otimes 4}} \ln \bigl( \mathbb{I} + t(\Psi_{12} + \Psi_{13} + \Psi_{14})\bigr)\right\}
\end{gather}
generates the maps studied in Section \ref{sec:Quartic} for $\lambda = 0$. Evidently $[\Psi_{1c}, \Psi_{1c'}]\neq0$ for $c\neq c'$ because they both act on the f\/irst copy of $V$. The observables are thus multi-traces of words in the alphabet $\{\Psi_{12}, \Psi_{13}, \Psi_{14}\}$.

It can be compared with the ${\rm O}(n)$ matrix model \cite{matrix}. It is a single matrix model for a mat\-rix~$\psi$ say of size $N^2\times N^2$ which acts on $V\otimes V$. Denote $\Psi_{12} = \psi \otimes \mathbb{I}_{V\otimes V}$ the matrix acting on $V\otimes V\otimes V\otimes V$ as $\psi$ on the f\/irst two factors and as the identity on the last two, and def\/ine $\Psi_{34} = \mathbb{I}_{V\otimes V} \otimes \psi$ acting non-trivially on the third and fourth factors. The matrix model is then
\begin{gather*}
\int d\psi\, \exp \big\{{-}N^2 \operatorname{Tr}_{V^{\otimes 2}} \psi^2 + n \operatorname{Tr}_{V^{\otimes 4}} \ln \bigl(\mathbb{I} + t(\Psi_{12} + \Psi_{34})\bigr)\big\}.
\end{gather*}
Remarkably, it has a similar form to \eqref{NecklaceMatrixModel}, except that the terms in the logarithm commute, $[\Psi_{12},\Psi_{34}] = 0$, because they act non-trivially on distinct copies of $V$. This leads to interpre\-ting~\eqref{NecklaceMatrixModel} as a non-commutative ${\rm O}(1)$ model.

It would therefore be interesting to directly write and solve the Schwinger--Dyson equations for~\eqref{NecklaceMatrixModel}, as done in the quartic melonic case in~\cite{IntermediateT4}. There are further interests in doing that. A~direct motivation is that it would in turn help to solve the Schwinger--Dyson equations for tensor models directly. This is a set of equations analogous to the Tutte/loop equations for colored triangulations in arbitrary dimensions. Those equations have been described in~\cite{BubbleAlgebra}. There were studied in the case of melonic bubbles in~\cite{SDE} at large~$N$, and it was shown in \cite{DoubleScaling} that the f\/irst corrections (colored graphs of degree $\omega(G) = d-2, d$) can be calculated from those equations. They have also been used for bubbles which result from the gluings of melonic bubbles with necklaces in the sense of Section \ref{sec:BubbleGluing}. The enumeration was performed using the Schwinger--Dyson equations in~\cite{MelonoPlanar} which reproduce the loop equations of multi-trace matrix models.

Schwinger--Dyson equations are also the key to go beyond the graphs which maximize the number of faces and to obtain a full classif\/ication of ${\mathcal G}(B)$ with respect to the number of faces. This is expected since this is the way it works in matrix models. Tutte's equations indeed give a recursion on the generating functions of maps of genus $g$ with $n$ marked faces which is the starting point of the topological recursion~\cite{TR}. Schwinger--Dyson equations also encode algebraic properties which in the case of combinatorial maps show integrability. In the case of tensor models, similar bilinear equations can be derived in the quartic melonic case~\cite{GiventalTensors}.

We therefore expect that the techniques presented here, together with a better understanding of the Schwinger--Dyson equations, will make it possible to identify more enhancements and to perform more enumeration of ${\mathcal G}_{\max}(G)$ thereby leading to a better exploration of the universality classes of colored triangulations, as well as a full classif\/ication and enumeration of ${\mathcal G}(B)$ for more bubbles $B$.

As often techniques are f\/irst developed for the purely combinatorial models and then adapted to more elaborate ones. This was the case with the discovery of melonic graphs as the graphs of vanishing degree, which then lead to using them to build renormalizable tensorial theories \cite{BGR+,BGR} and group f\/ield theories \cite{MelonicGFT,MelonicGFT++,MelonicGFT+}. The bijection we have presented here was even used in the quartic case for such renormalizable tensorial theories to get closed equations on some Green functions \cite{LahocheOritiRivasseau} (this is the equivalent of enumerating for renormalizable models).

Hopefully, our techniques could also be adapted to more general families of triangulations, such as the multi-orientable one in $d=3$ \cite{MO} and the family introduced in \cite{RealTensors} which contain the multi-orientable and colored triangulations in $d=3$.

\subsection*{Acknowledgements}

This research was supported by the ANR MetACOnc project ANR-15-CE40-0014.

\pdfbookmark[1]{References}{ref}
\LastPageEnding


\begin{thebibliography}{99}
\footnotesize\itemsep=0pt

\bibitem{AlvarezBarbon}
Alvarez-Gaum{\'e} L., Barb{\'o}n J.L.F., Crnkovi{\'c} {\v{C}}., A proposal for
 strings at {$D>1$}, \href{http://dx.doi.org/10.1016/0550-3213(93)90020-P}{\textit{Nuclear Phys.~B}} \textbf{394} (1993), 383--422,
 \href{http://arxiv.org/abs/hep-th/9208026}{hep-th/9208026}.

\bibitem{AmbjornTensors}
Ambj{\o}rn J., Durhuus B., J{\'o}nsson T., Three-dimensional simplicial quantum
 gravity and generalized matrix models, \href{http://dx.doi.org/10.1142/S0217732391001184}{\textit{Modern Phys. Lett.~A}}
 \textbf{6} (1991), 1133--1146.

\bibitem{MelonicGFT}
Baratin A., Carrozza S., Oriti D., Ryan J., Smerlak M., Melonic phase
 transition in group f\/ield theory, \href{http://dx.doi.org/10.1007/s11005-014-0699-9}{\textit{Lett. Math. Phys.}} \textbf{104}
 (2014), 1003--1017, \href{http://arxiv.org/abs/1307.5026}{arXiv:1307.5026}.

\bibitem{BarbonDemeterfi}
Barb{\'o}n J.L.F., Demeterf\/i K., Klebanov I.R., Schmidhuber C., Correlation
 functions in matrix models modif\/ied by wormhole terms, \href{http://dx.doi.org/10.1016/0550-3213(95)00084-6}{\textit{Nuclear
 Phys.~B}} \textbf{440} (1995), 189--214, \href{http://arxiv.org/abs/hep-th/9501058}{hep-th/9501058}.

\bibitem{BGR+}
Ben~Geloun J., Two- and four-loop {$\beta$}-functions of rank-4 renormalizable
 tensor f\/ield theories, \href{http://dx.doi.org/10.1088/0264-9381/29/23/235011}{\textit{Classical Quantum Gravity}} \textbf{29} (2012),
 235011, 40~pages, \href{http://arxiv.org/abs/1205.5513}{arXiv:1205.5513}.

\bibitem{BubbleCounting}
Ben~Geloun J., Ramgoolam S., Counting tensor model observables and branched
 covers of the 2-sphere, \href{http://dx.doi.org/10.4171/AIHPD/4}{\textit{Ann. Inst. Henri Poincar\'e~D}} \textbf{1}
 (2014), 77--138, \href{http://arxiv.org/abs/1307.6490}{arXiv:1307.6490}.

\bibitem{BGR}
Ben~Geloun J., Rivasseau V., A renormalizable 4-dimensional tensor f\/ield
 theory, \href{http://dx.doi.org/10.1007/s00220-012-1549-1}{\textit{Comm. Math. Phys.}} \textbf{318} (2013), 69--109,
 \href{http://arxiv.org/abs/1111.4997}{arXiv:1111.4997}.

\bibitem{New1/N}
Bonzom V., New {$1/N$} expansions in random tensor models, \href{http://dx.doi.org/10.1007/JHEP06(2013)062}{\textit{J.~High
 Energy Phys.}} \textbf{2013} (2013), no.~6, 062, 25~pages, \href{http://arxiv.org/abs/1211.1657}{arXiv:1211.1657}.

\bibitem{SDE}
Bonzom V., Revisiting random tensor models at large {$N$} via the
 {S}chwinger--{D}yson equations, \href{http://dx.doi.org/10.1007/JHEP03(2013)160}{\textit{J.~High Energy Phys.}} \textbf{2013}
 (2013), no.~3, 160, 25~pages, \href{http://arxiv.org/abs/1208.6216}{arXiv:1208.6216}.

\bibitem{MelonoPlanar}
Bonzom V., Delepouve T., Rivasseau V., Enhancing non-melonic triangulations: a
 tensor model mixing melonic and planar maps, \href{http://dx.doi.org/10.1016/j.nuclphysb.2015.04.004}{\textit{Nuclear Phys.~B}}
 \textbf{895} (2015), 161--191, \href{http://arxiv.org/abs/1502.01365}{arXiv:1502.01365}.

\bibitem{ColoredMelonic}
Bonzom V., Gurau R., Riello A., Rivasseau V., Critical behavior of colored
 tensor models in the large~{$N$} limit, \href{http://dx.doi.org/10.1016/j.nuclphysb.2011.07.022}{\textit{Nuclear Phys.~B}} \textbf{853}
 (2011), 174--195, \href{http://arxiv.org/abs/1105.3122}{arXiv:1105.3122}.

\bibitem{Uncoloring}
Bonzom V., Gurau R., Rivasseau V., Random tensor models in the large $N$ limit:
 uncoloring the colored tensor models, \href{http://dx.doi.org/10.1103/PhysRevD.85.084037}{\textit{Phys. Rev.~D}} \textbf{85}
 (2012), 084037, 12~pages, \href{http://arxiv.org/abs/1202.3637}{arXiv:1202.3637}.

\bibitem{DoubleScaling}
Bonzom V., Gurau R., Ryan J.P., Tanasa A., The double scaling limit of random
 tensor models, \href{http://dx.doi.org/10.1007/JHEP09(2014)051}{\textit{J.~High Energy Phys.}} \textbf{2014} (2014), no.~9,
 051, 49~pages, \href{http://arxiv.org/abs/1404.7517}{arXiv:1404.7517}.

\bibitem{StuffedWalshMaps}
Bonzom V., Lionni L., Rivasseau V., Colored triangulations of arbitrary
 dimensions are stuf\/fed {W}alsh maps, \href{http://arxiv.org/abs/1508.03805}{arXiv:1508.03805}.

\bibitem{MelonicGFT++}
Carrozza S., Oriti D., Rivasseau V., Renormalization of a {${\rm SU}(2)$}
 tensorial group f\/ield theory in three dimensions, \href{http://dx.doi.org/10.1007/s00220-014-1928-x}{\textit{Comm. Math. Phys.}}
 \textbf{330} (2014), 581--637, \href{http://arxiv.org/abs/1303.6772}{arXiv:1303.6772}.

\bibitem{MelonicGFT+}
Carrozza S., Oriti D., Rivasseau V., Renormalization of tensorial group f\/ield
 theories: {A}belian {${\rm U}(1)$} models in four dimensions, \href{http://dx.doi.org/10.1007/s00220-014-1954-8}{\textit{Comm.
 Math. Phys.}} \textbf{327} (2014), 603--641, \href{http://arxiv.org/abs/1207.6734}{arXiv:1207.6734}.

\bibitem{RealTensors}
Carrozza S., Tanasa A., ${\rm O}(N)$ random tensor models, \href{http://arxiv.org/abs/1512.06718}{arXiv:1512.06718}.

\bibitem{GiventalTensors}
Dartois S., A {G}ivental-like formula and bilinear identities for tensor
 models, \href{http://dx.doi.org/10.1007/JHEP08(2015)129}{\textit{J.~High Energy Phys.}} \textbf{2015} (2015), no.~8, 129,
 19~pages, \href{http://arxiv.org/abs/1409.5621}{arXiv:1409.5621}.

\bibitem{Das}
Das S.R., Dhar A., Sengupta A.M., Wadia S.R., New critical behavior in {$d=0$}
 large-{$N$} matrix models, \href{http://dx.doi.org/10.1142/S0217732390001165}{\textit{Modern Phys. Lett.~A}} \textbf{5} (1990),
 1041--1056.

\bibitem{BorelQuartic}
Delepouve T., Gurau R., Rivasseau V., Universality and {B}orel summability of
 arbitrary quartic tensor models, \href{http://dx.doi.org/10.1214/14-AIHP655}{\textit{Ann. Inst. Henri Poincar\'e Probab.
 Stat.}} \textbf{52} (2016), 821--848, \href{http://arxiv.org/abs/1403.0170}{arXiv:1403.0170}.

\bibitem{matrix}
Di~Francesco P., Ginsparg P., Zinn-Justin J., {$2$}{D} gravity and random
 matrices, \href{http://dx.doi.org/10.1016/0370-1573(94)00084-G}{\textit{Phys. Rep.}} \textbf{254} (1995), 133,
 \href{http://arxiv.org/abs/hep-th/9306153}{hep-th/9306153}.

\bibitem{TR}
Eynard B., Topological expansion for the 1-{H}ermitian matrix model correlation
 functions, \href{http://dx.doi.org/10.1088/1126-6708/2004/11/031}{\textit{J.~High Energy Phys.}} \textbf{2004} (2004), no.~11, 031,
 35~pages, \href{http://arxiv.org/abs/hep-th/0407261}{hep-th/0407261}.

\bibitem{FerriGagliardi}
Ferri M., Gagliardi C., Crystallisation moves, \href{http://dx.doi.org/10.2140/pjm.1982.100.85}{\textit{Pacific~J. Math.}}
 \textbf{100} (1982), 85--103.

\bibitem{AC}
Flajolet P., Sedgewick R., Analytic combinatorics, \href{http://dx.doi.org/10.1017/CBO9780511801655}{Cambridge University Press},
 Cambridge, 2009.

\bibitem{GouldenJacksonBook}
Goulden I.P., Jackson D.M., Combinatorial enumeration, \textit{A Wiley-Interscience
 Publication}, John Wiley \& Sons, Inc., New York, 1983.

\bibitem{GrossTensors}
Gross M., Tensor models and simplicial quantum gravity in {$>2-D$}, \href{http://dx.doi.org/10.1016/S0920-5632(05)80015-5}{\textit{Nuclear Phys.~B Proc. Suppl.}} \textbf{25A} (1992), 144--149.

\bibitem{Lost}
Gurau R., Lost in translation: topological singularities in group f\/ield theory,
 \href{http://dx.doi.org/10.1088/0264-9381/27/23/235023}{\textit{Classical Quantum Gravity}} \textbf{27} (2010), 235023, 20~pages,
 \href{http://arxiv.org/abs/1006.0714}{arXiv:1006.0714}.

\bibitem{LargeN1}
Gurau R., The {$1/N$} expansion of colored tensor models, \href{http://dx.doi.org/10.1007/s00023-011-0101-8}{\textit{Ann. Henri
 Poincar\'e}} \textbf{12} (2011), 829--847, \href{http://arxiv.org/abs/1011.2726}{arXiv:1011.2726}.

\bibitem{TreeAlgebra}
Gurau R., A generalization of the {V}irasoro algebra to arbitrary dimensions,
 \href{http://dx.doi.org/10.1016/j.nuclphysb.2011.07.009}{\textit{Nuclear Phys.~B}} \textbf{852} (2011), 592--614, \href{http://arxiv.org/abs/1105.6072}{arXiv:1105.6072}.

\bibitem{Lost++}
Gurau R., Reply to comment on `{L}ost in translation: topological singularities
 in group f\/ield theory', \href{http://dx.doi.org/10.1088/0264-9381/28/17/178002}{\textit{Classical Quantum Gravity}} \textbf{28}
 (2011), 178002, 2~pages, \href{http://arxiv.org/abs/1108.4966}{arXiv:1108.4966}.

\bibitem{LargeN3}
Gurau R., The complete {$1/N$} expansion of colored tensor models in arbitrary
 dimension, \href{http://dx.doi.org/10.1007/s00023-011-0118-z}{\textit{Ann. Henri Poincar\'e}} \textbf{13} (2012), 399--423,
 \href{http://arxiv.org/abs/1102.5759}{arXiv:1102.5759}.

\bibitem{BubbleAlgebra}
Gurau R., The {S}chwinger--{D}yson equations and the algebra of constraints of
 random tensor models at all orders, \href{http://dx.doi.org/10.1016/j.nuclphysb.2012.07.028}{\textit{Nuclear Phys.~B}} \textbf{865}
 (2012), 133--147, \href{http://arxiv.org/abs/1203.4965}{arXiv:1203.4965}.

\bibitem{BeyondPerturbation}
Gurau R., The {$1/N$} expansion of tensor models beyond perturbation theory,
 \href{http://dx.doi.org/10.1007/s00220-014-1907-2}{\textit{Comm. Math. Phys.}} \textbf{330} (2014), 973--1019,
 \href{http://arxiv.org/abs/1304.2666}{arXiv:1304.2666}.

\bibitem{Universality}
Gurau R., Universality for random tensors, \href{http://dx.doi.org/10.1214/13-AIHP567}{\textit{Ann. Inst. Henri Poincar\'e
 Probab. Stat.}} \textbf{50} (2014), 1474--1525, \href{http://arxiv.org/abs/1111.0519}{arXiv:1111.0519}.

\bibitem{GurauKrajewski}
Gurau R., Krajewski T., Analyticity results for the cumulants in a random
 matrix model, \href{http://dx.doi.org/10.4171/AIHPD/17}{\textit{Ann. Inst. Henri Poincar\'e D}} \textbf{2} (2015),
 169--228, \href{http://arxiv.org/abs/1409.1705}{arXiv:1409.1705}.

\bibitem{LargeN2}
Gurau R., Rivasseau V., The $1/N$ expansion of colored tensor models in
 arbitrary dimension, \href{http://dx.doi.org/10.1209/0295-5075/95/50004}{\textit{Europhys. Lett.}} \textbf{95} (2011), 50004,
 5~pages, \href{http://arxiv.org/abs/1101.4182}{arXiv:1101.4182}.

\bibitem{GurauRyanReview}
Gurau R., Ryan J.P., Colored tensor models~-- a review, \href{http://dx.doi.org/10.3842/SIGMA.2012.020}{\textit{SIGMA}}
 \textbf{8} (2012), 020, 78~pages, \href{http://arxiv.org/abs/1109.4812}{arXiv:1109.4812}.

\bibitem{BP}
Gurau R., Ryan J.P., Melons are branched polymers, \href{http://dx.doi.org/10.1007/s00023-013-0291-3}{\textit{Ann. Henri
 Poincar\'e}} \textbf{15} (2014), 2085--2131, \href{http://arxiv.org/abs/1302.4386}{arXiv:1302.4386}.

\bibitem{GurauSchaeffer}
Gurau R., Schaef\/fer G., Regular colored graphs of positive degree,
 \href{http://arxiv.org/abs/1307.5279}{arXiv:1307.5279}.

\bibitem{NLO}
Kaminski W., Oriti D., Ryan J.P., Towards a double-scaling limit for tensor
 models: probing sub-dominant orders, \href{http://dx.doi.org/10.1088/1367-2630/16/6/063048}{\textit{New~J. Phys.}} \textbf{16}
 (2014), 063048, 36~pages, \href{http://arxiv.org/abs/1304.6934}{arXiv:1304.6934}.

\bibitem{KlebanovHashimoto}
Klebanov I.R., Hashimoto A., Non-perturbative solution of matrix models
 modif\/ied by trace-squared terms, \href{http://dx.doi.org/10.1016/0550-3213(94)00518-J}{\textit{Nuclear Phys.~B}} \textbf{434}
 (1995), 264--282, \href{http://arxiv.org/abs/hep-th/9409064}{hep-th/9409064}.

\bibitem{Korchemsky}
Korchemsky G.P., Loops in the curvature matrix model, \href{http://dx.doi.org/10.1016/0370-2693(92)91328-7}{\textit{Phys. Lett.~B}}
 \textbf{296} (1992), 323--334, \href{http://arxiv.org/abs/hep-th/9206088}{hep-th/9206088}.

\bibitem{LahocheOritiRivasseau}
Lahoche V., Oriti D., Rivasseau V., Renormalization of an {A}belian tensor
 group f\/ield theory: solution at leading order, \href{http://dx.doi.org/10.1007/JHEP04(2015)095}{\textit{J.~High Energy Phys.}}
 \textbf{2015} (2015), no.~4, 095, 41~pages, \href{http://arxiv.org/abs/1501.02086}{arXiv:1501.02086}.

\bibitem{Lins}
Lins S., Gems, computers and attractors for {$3$}-manifolds, \href{http://dx.doi.org/10.1142/9789812796196}{\textit{Series on
 Knots and Everything}}, Vol.~5, World Scientif\/ic Publishing Co., Inc., River
 Edge, NJ, 1995.

\bibitem{IntermediateT4}
Nguyen V.A., Dartois S., Eynard B., An analysis of the intermediate f\/ield
 theory of $T^4$ tensor model, \href{http://dx.doi.org/10.1007/JHEP01(2015)013}{\textit{J.~High Energy Phys.}} \textbf{2015}
 (2015), no.~1, 013, 17~pages, \href{http://arxiv.org/abs/1409.5751}{arXiv:1409.5751}.

\bibitem{SasakuraTensors}
Sasakura N., Tensor model for gravity and orientability of manifold,
 \href{http://dx.doi.org/10.1142/S0217732391003055}{\textit{Modern Phys. Lett.~A}} \textbf{6} (1991), 2613--2623.

\bibitem{Lost+}
Smerlak M., Comment on `{L}ost in translation: topological singularities in
 group f\/ield theory', \href{http://dx.doi.org/10.1088/0264-9381/28/17/178001}{\textit{Classical Quantum Gravity}} \textbf{28} (2011),
 178001, 3~pages, \href{http://arxiv.org/abs/1102.1844}{arXiv:1102.1844}.

\bibitem{MO}
Tanasa A., The multi-orientable random tensor model, a review, \href{http://dx.doi.org/10.3842/SIGMA.2016.056}{\textit{SIGMA}}
 \textbf{12} (2016), 056, 23~pages, \href{http://arxiv.org/abs/1512.02087}{arXiv:1512.02087}.

\bibitem{Walsh}
Walsh T.R.S., Hypermaps versus bipartite maps, \href{http://dx.doi.org/10.1016/0095-8956(75)90042-8}{\textit{J.~Combinatorial Theory
 Ser.~B}} \textbf{18} (1975), 155--163.

\end{thebibliography}
\end{document}